\documentclass[prb,aps,amsmath,amssymb,amsfonts,floatfix,superscriptaddress,showpacs,twocolumn]{revtex4}
\usepackage{graphicx}
\usepackage{color}
\usepackage{bbm}
\usepackage[utf8]{inputenc}
\usepackage{dcolumn}
\usepackage{hyperref}
\hypersetup{colorlinks=true,breaklinks,linkcolor=black,urlcolor=blue,citecolor=blue} 
                                                                                                                                                                                                                                                                                                 
\newcommand{\kv}{\ensuremath{\mathbf{k}}}                                                                                                                                          
                                                                                                                                  
\newcommand{\T}{\ensuremath{T_{\tau}}}                                                                                                                                                    
\newcommand{\dtau}{{\ensuremath{\partial_\tau}}}                                                                                                                                   
\newcommand{\iom}{{\ensuremath{i\nu}}}                                                                                                                                          
\newcommand{\av}[1]{\ensuremath{\langle #1 \rangle}}                                                                                                                               
                                                                                                                                         
\newcommand\Let{\mathrel{\mathop:\!\!=}}

\def \sgn {\mathop {\rm sgn}}                                                                                                                                                      
\def \Im {\mathop {\rm Im}}

\begin{document}

\title{Improved Estimators for the Self-Energy and Vertex Function
in Hybridization Expansion Continuous-Time Quantum Monte Carlo Simulations}

\author{Hartmut Hafermann}
\affiliation{Centre de Physique Th\'eorique, \'Ecole Polytechnique, CNRS, 91128 Palaiseau Cedex, France}
\author{Kelly R. Patton}
\affiliation{Department of Physics and Astronomy, Louisiana State University, Baton Rouge, Louisiana 70803}
\author{Philipp Werner}
\affiliation{Theoretische Physik, ETH Zurich, 8093 Z\"urich, Switzerland}

\date{\today}

\begin{abstract}

We propose efficient measurement procedures for the self-energy and vertex function of the Anderson impurity model within the hybridization expansion continuous-time quantum Monte Carlo algorithm. The method is based on the measurement of higher-order correlation functions related to the quantities being sought through the equation of motion, a technique previously introduced in the NRG context. For the case of interactions of density-density type, the additional correlators can be obtained at essentially no additional computational cost.
In combination with a recently introduced method for filtering the Monte Carlo noise using a representation in terms of orthogonal polynomials, we obtain data with unprecedented accuracy. 
This leads to an enhanced stability in analytical continuations of the self-energy or in two-particle based theories such as the dual fermion approach.
As an illustration of the method we reexamine the previously reported spin-freezing and high-spin to low-spin transitions in a two-orbital model with density-density interactions. In both cases, the vertex function undergoes significant changes, which suggests significant corrections to the dynamical mean-field solutions in dual fermion calculations.

\end{abstract}

\pacs{
71.10.-w,
71.27.+a,
71.30.+h
}

\maketitle

\section{Introduction}

Continuous-time quantum Monte Carlo solvers (for a recent review, see Ref.~\onlinecite{CTQMCRMP}) have become an important tool for the study of the Anderson impurity model (AIM) and its multi-orbital generalizations, due to their accuracy, efficiency and ability to treat arbitrary interaction terms. While the AIM plays a fundamental role in various areas of condensed matter physics, the continuous-time solvers have been developed and primarily applied in the context of dynamical mean-field theory (DMFT),\cite{Georges96} which maps correlated lattice models to an appropriately defined quantum impurity model. 

Impurity models with multiple orbitals are important for two different reasons:
(i) The combination of density functional theory and DMFT \cite{KotliarRMP06} provides a formalism for the calculation and prediction of properties of strongly correlated materials. The description of materials with open d- or f-shells requires the solution of impurity models with up to five or seven correlated orbitals, respectively.
(ii) In the context of cluster extensions of DMFT,\cite{MaierReview} the lattice is mapped onto a cluster of impurities in order to account for spatial correlations, and it is desirable to solve clusters with as many sites as possible, in order to be able to infer reliable predictions for the infinite system.
For cluster DMFT calculations of simple models, the interaction expansion or weak-coupling algorithm, which is based on an expansion of the partition function in the interaction (henceforth referred to as CT-INT\cite{Rubtsov05})  and the related continuous-time auxiliary field algorithm (CT-AUX\cite{CTAUX}) have proven useful.\cite{CTAUXapp1,CTAUXapp2} Here the number of interaction terms and hence the perturbation order grow linearly with the number of cluster sites. In the context of \emph{ab initio} calculations of correlated materials, however, the number of interaction terms grows at least as the square of the number of orbitals and hence the hybridization expansion algorithm \cite{Werner06,WernerPRBLong,Haule07} (abbreviated CT-HYB) is the method of choice.

For the latter, the calculation of the one-electron self-energy has proven problematic.
The calculation from Dyson's equation, where it is evaluated as the difference between the inverses of two Green's functions, is highly susceptible to noise in the numerical data. In contrast to CT-INT, the Green's function in CT-HYB is not measured as a correction to a known function (the noninteracting Green's function $G_{0}$), which leads to large noise in the intermediate to high-frequency region. Except for the low-frequency region, better statistics is needed for the calculation of the self-energy in CT-HYB to obtain results of comparable accuracy as in CT-INT.\cite{Gull07}

Similar problems arise in the calculation of the impurity vertex function. While the vertex allows one to access response functions of a system, interest in this quantity has recently grown in particular due to the advent of novel diagrammatic extensions of the dynamical mean-field theory.\cite{kusunose,toschi,multiscale,dualfermion1,ldfa} In these approaches, spatial correlations beyond DMFT are included through two-particle field theories, which involve the two-particle irreducible (in the dynamical vertex approximation \cite{toschi}) or reducible (in the dual fermion approach \cite{dualfermion1,ldfa}) vertex function.
Thus far these approaches mainly relied on exact diagonalization (ED) or CT-INT for the calculation of the impurity vertex functions. Only few applications employing the CT-HYB algorithm for measuring the two-particle function have been reported because of the aforementioned problems. The measurement of frequency dependent two-particle functions is particularly challenging for multiorbital models. For example, in a recent letter by Park \emph{et al.}, this problem was circumvented by neglecting the frequency dependence of the irreducible vertex function and hence setting it to a constant.\cite{ParkPRL}

On the other hand, the CT-HYB approach with its low perturbation order and favorable sign statistics\cite{Gull07} appears to be the most suitable, in principle, for measuring vertex functions in realistic multiorbital models.

In this paper, we propose an efficient method for calculating the self-energy and vertex function which eliminates the noise problems. Through the equation of motion, the self-energy and vertex function are related to higher-order correlation functions, which can be measured in CT-HYB. In combination with a recently developed method to eliminate the Monte Carlo noise through a representation in terms of orthogonal polynomials,\cite{legendre} the present approach yields considerably more accurate results compared to the standard measurements.

\section{Method}
\subsection{Model}

\begin{figure}[t]
\includegraphics[scale=0.5,angle=0]{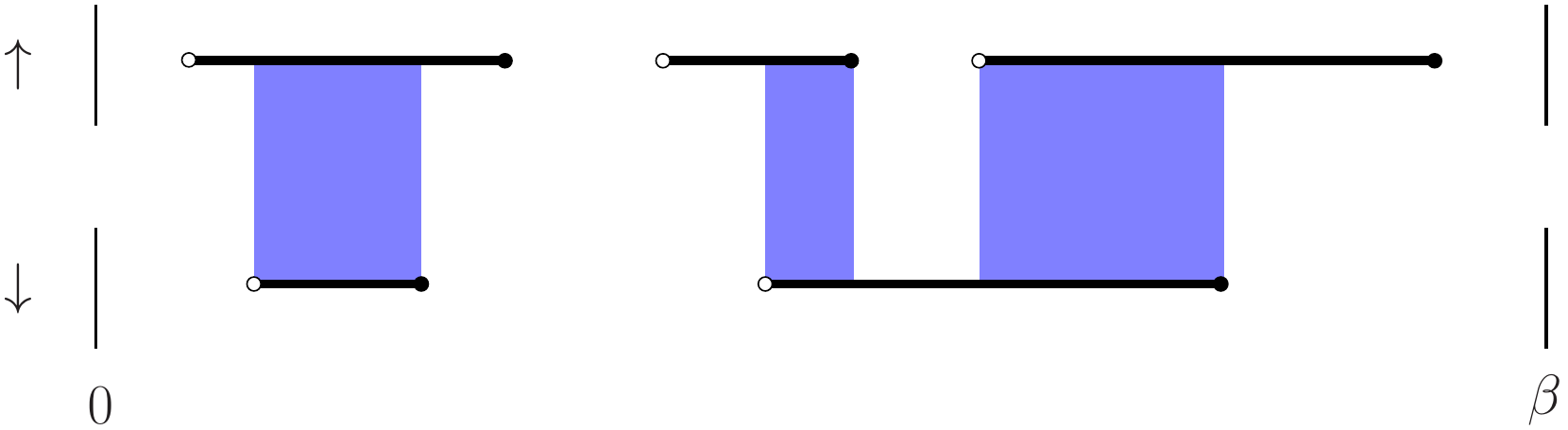}
\caption{\label{fig:overlap} (Color online) Typical segment configuration in the hybridization expansion continuous-time quantum Monte Carlo simulation of the one-orbital AIM. Each segment marks an imaginary-time interval in which an electron of given spin resides on the impurity. The overlap between the up- and down-spin segments (blue or gray shaded area) determines the interaction energy.}
\end{figure}

We consider the single-site, multiorbital AIM with interactions of density-density type, described by the Hamiltonian
\begin{align}
H_\text{AIM} = \sum_{\kv i} \varepsilon_\kv^i f_{\kv i}^\dagger f_{\kv i} &+ \sum_i \varepsilon_i n_i+ \frac{1}{2}\sum_{ij} U_{ij} n_i n_j \nonumber\\	
&+ 
\sum_{\kv ij}\left(c_i^\dagger V_\kv^{ij} f_{\kv j} + f^\dagger_{\kv i} V_\kv^{* ij} c_j \right),
\label{eqn:hamiltonian}
\end{align}
where latin subscripts denote a `flavor' index, i.e. a combined index for spin- and orbital degrees of freedom. The operators $c_{i}^{\dagger}$ $(c_{i})$ create (destroy) an electron with flavor $i$ on the impurity site, while $f_{\kv i}^{\dagger}$ $(f_{\kv i})$ creates (destroys) a conduction band electron in band $i$ with momentum $\kv$ ($\varepsilon_\kv^i $ is the corresponding dispersion). The impurity levels are denoted by $\varepsilon_{i}$ and the matrix $U_{ij}$ contains the various interaction parameters for the interactions of density-density type. Electrons with band-flavor index $j$ and momentum $\kv$ are allowed to couple to electrons with any other flavor index $i$ as described by the hybridization matrix $V_\kv^{ij}$.
Integrating out the noninteracting conduction band electrons gives rise to the hybridization function
\begin{align}
\Delta_{ab}(\iom) = \sum_{\kv j}\frac{V_\kv^{aj} V_\kv^{* jb}}{\iom-\varepsilon_\kv^j}.
\label{eqn:hybridization}
\end{align}

Here we use an implementation based on the segment picture of the hybridization expansion algorithm.\cite{Werner06} 
The segment picture applies whenever operators of a given flavor appear in alternating order, which is always the case for an interaction of density-density type.

 In this case the trace over the sequence of impurity creation and annihilation operators (which defines the Monte Carlo configuration) does not have to be computed explicitly. In order to evaluate the ratio of traces one instead only needs to compute the length of the segments for the different flavors and their overlaps. 
This yields a very efficient algorithm for multiorbital problems with density-density interaction, which --- as long as the determinant calculation dominates the computational effort --- scales linearly in the number of flavors.\cite{CTQMCRMP} 
A possible segment configuration for the one-orbital AIM is illustrated in Fig.~\ref{fig:overlap}. The segments represent imaginary-time intervals in which an electron of given flavor (spin) resides on the impurity. Overlapping segments correspond to time intervals in which the impurity is doubly occupied.
As we will see below, the additional higher-order correlation functions which arise in the expressions for the self-energy and vertex function in the improved estimator can be measured at essentially no additional computational cost in this segment representation. 
The generalization to the case of interactions of non-density-density type, such as spin-flip and pair hopping terms, is in principle straightforward. The measurement of the required correlation functions in the hybridization expansion algorithm, however, becomes more involved.

In the following subsection, we first introduce and illustrate the procedure for the self-energy and consider the generalization to the vertex function in a second separate subsection.

\subsection{Self-energy}

\begin{figure}[t]
\includegraphics[scale=0.225,angle=0]{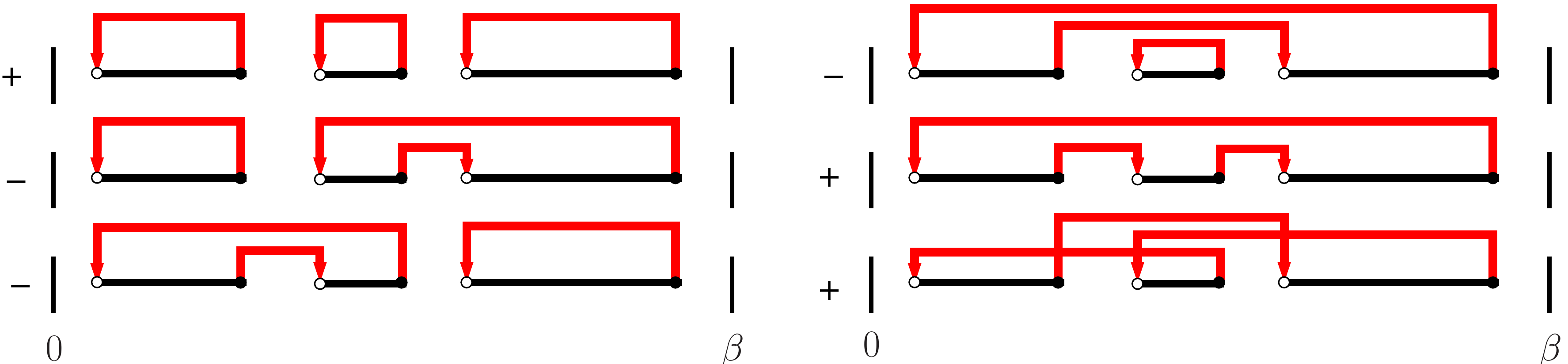} 
\caption{\label{fig:scconfig} (Color online) The $3!$ ways of connecting hybridization lines in a configuration with 3 segments for a timeline of a given flavor. The sign of each diagram is indicated. For non-diagonal hybridization, the lines also connect segments with different flavors.}
\end{figure}

The asymptotic tail of the self-energy can be obtained from a high-frequency expansion, which requires measurements of single- and two-particle equal-time correlators (see e.g. Ref. \onlinecite{sigmatails} and references therein). This eliminates the noise at high frequencies. However, the choice of the cutoff value cannot easily be automatized, and as we will show (see, e.g., the inset of Fig. \ref{fig:sigma_legendre}), the noise problem of the standard measurement is significant even at frequencies for which the self-energy clearly has not yet reached its asymptotic behavior.

The method presented here yields accurate results over the entire frequency range and considerably reduces the noise at intermediate to high frequencies. 
An alternative approach to reduce the Monte Carlo noise is to measure observables in an orthogonal polynomial representation.\cite{legendre} This corresponds to an advantageous change of basis, which yields a compact representation of observables and allows one to filter the Monte Carlo noise, without any loss of information. We note that such a procedure does not reduce the required statistics and hence for optimal results the two approaches should be combined.

The computation of the self-energy from higher-order correlation functions of the impurity model has been introduced previously in the context of numerical renormalization group (NRG) \cite{NRG} calculations.
Here we provide the expression for the multiorbital case and demonstrate the usefulness of this scheme for the CT-HYB algorithm.
An expression for the self-energy in terms of a two-particle correlator is obtained from the equation of motion for the Green's function. One may write the equation of motion in terms of a derivative with respect to either the first $(\tau)$ or second $(\tau')$ of its arguments. For reasons outlined in the appendix, we choose the second option. Here we only present a brief outline of the derivation. Details can be found in Appendix \ref{app:sigma}.

\begin{figure}[b]
\includegraphics[scale=0.475,angle=0]{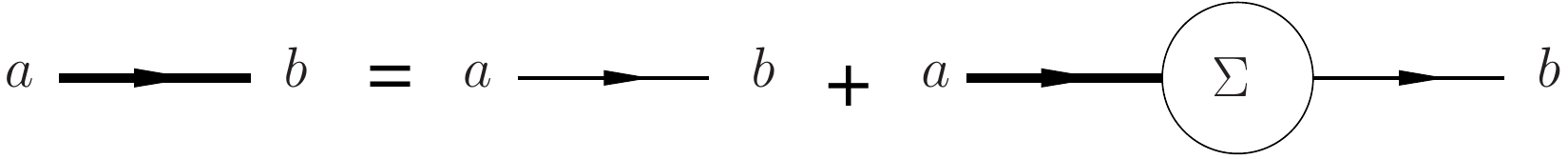} 
\caption{\label{fig:sigma} Diagrammatical representation of the Green's function $G_{ab}$ in terms the self energy $\Sigma$, Eq. \eqref{eqn:dyson1}. Thick lines denote fully dressed propagators, thin lines denote bare propagators.
}
\end{figure}

The time derivative of the Green's function 
\begin{equation}
G_{ab}(\tau-\tau')\Let - \av{\T c_a(\tau) c_b^\dagger(\tau')}
\end{equation}
is given by
\begin{align}
\partial_{\tau'}  G_{ab}(\tau-\tau') = \delta(\tau-\tau')\delta_{ab} - \av{\T c_a(\tau) \dtau_{'} c_b^\dagger(\tau')},
\label{eqn:Gderiv}
\end{align}
where $\T$ is the usual time-ordering operator. Its equation of motion follows from the one for the operator $c^{\dagger}_{b}$ taken in the Heisenberg representation:
\begin{align}
\dtau_{'} c_b^{\dagger}(\tau') = [H,c_b^{\dagger}](\tau').
\label{eqn:eqnofmotioncdag}
\end{align}
We introduce the following correlation functions:
\begin{align}
G_{\kv ab}^\text{cf}(\tau-\tau') &\Let -\av{\T c_{a}(\tau) f^\dagger_{\kv b}(\tau')},\\
F^{j}_{ab}(\tau-\tau') &\Let -\av{\T c_a(\tau) c^\dagger_b(\tau')n_j(\tau') },
\end{align}
together with their Fourier transforms $G_{\kv ab}^\text{cf}(\iom)$ and $F^{j}_{ab}(\iom)$.
With an application in the context of DMFT in mind, we furthermore introduce the noninteracting Green's function of the impurity model,
\begin{equation}
G_{0\, ab}^{-1}(\iom) = (\iom -\varepsilon_b)\delta_{ab} - \Delta_{ab}(\iom),
\end{equation}
where $\Delta_{ab}(\iom)$ is the hybridization function.
Evaluating the commutator in Eq.~\eqref{eqn:eqnofmotioncdag} with the Hamiltonian \eqref{eqn:hamiltonian} yields
\begin{align}
[H,c_b^{\dagger}]=\varepsilon_b c_b^{\dagger} + \frac{1}{2} \sum_{j} (U_{jb}+ U_{bj}) c_{b}^{\dagger}n_j  + \sum_{\kv j} f_{\kv j}^{\dagger} V_\kv^{*\, jb},
\end{align}
so that Eq.~\eqref{eqn:Gderiv} in Fourier space can be written
\begin{align}
G_{ab}(\iom) = G_{0,ab}(\iom) &- \sum_{ij} G_{ai}(\iom) \Delta_{ij}(\iom) G_{0,jb}(\iom)\nonumber\\
&+ \sum_{\kv ij} G_{\kv ai}^\text{fc}(\iom)  V_\kv^{*\,ij} G_{0,jb}(\iom) \nonumber\\
&+ \frac{1}{2}\sum_{ij} (U_{ji}+U_{ij}) F^{j}_{ai}(\iom)G_{0,ib}(\iom).
\label{eqn:eqnofmotiongen2dag}
\end{align}
The function $G_{\kv ab}^\text{cf}(\iom)$ in turn can be eliminated by expressing it in terms of the impurity Green's function through its equation of motion. In Fourier space, it reads
\begin{equation}
G_{\kv ab}^\text{cf}(\iom) = \sum_i G_{ai}(\iom)\frac{V_\kv^{ib}}{\iom - \varepsilon_\kv^{b}}.
\end{equation}
Inserting this into \eqref{eqn:eqnofmotiongen2dag} and using the expression for the hybridization function \eqref{eqn:hybridization}, the hybridization terms cancel. Comparing the resulting expression with Dyson's equation (see Fig. \ref{fig:sigma}), 
\begin{equation}
G_{ab}(\iom) = G_{0,ab}(\iom) + \sum_{ij} G_{ai}(\iom) \Sigma_{ij}(\iom) G_{0,jb}
(\iom) ,
\label{eqn:dyson1}
\end{equation}
we see that the impurity self-energy can be expressed in the following form:
\begin{equation}
\Sigma_{ab}(\iom) = \frac{1}{2}\sum_{ij} G^{-1}_{ai}(\iom) (U_{jb}+U_{bj}) F^{j}_{ib}(\iom),
\label{eqn:sigmafinal}
\end{equation}
which for a single orbital model reduces to the result given in Ref. \onlinecite{NRG}.
This equation relates the self-energy to two measurable quantities, the impurity Green's function $G_{ab}(\iom)$ and an additional correlation function $F_{ab}^{j}(\iom)$.

In order to show how correlation functions are measured in the CT-HYB, we recall that in this algorithm, one samples over configurations whose weight is given by a determinant of hybridization functions times a trace over a sequence of operators. 
In the segment picture, a configuration $\mathcal{C}$ is visualized by a collection of segments on the timeline from $0$ to $\beta$ for each flavor ($\beta$ is the inverse temperature). 
A typical configuration for a single-orbital model with Hubbard interaction $Un_{\uparrow}n_{\downarrow}$ is depicted in Fig. \ref{fig:overlap}. 
The start of a segment (open circles) is associated with an impurity creation operator, while an impurity annihilation operator is associated with the segment endpoint (closed circles). A segment hence marks the imaginary time interval in which the impurity is occupied by an electron of a given flavor. 
The segments are connected by hybridization lines in all possible ways, which is the interpretation of the determinant (see Fig. \ref{fig:scconfig}). For the case of non-diagonal hybridization considered here, the segments are also connected by hybridization lines linking different flavors (not shown).

\begin{figure}[t]
\includegraphics[scale=0.675,angle=0]{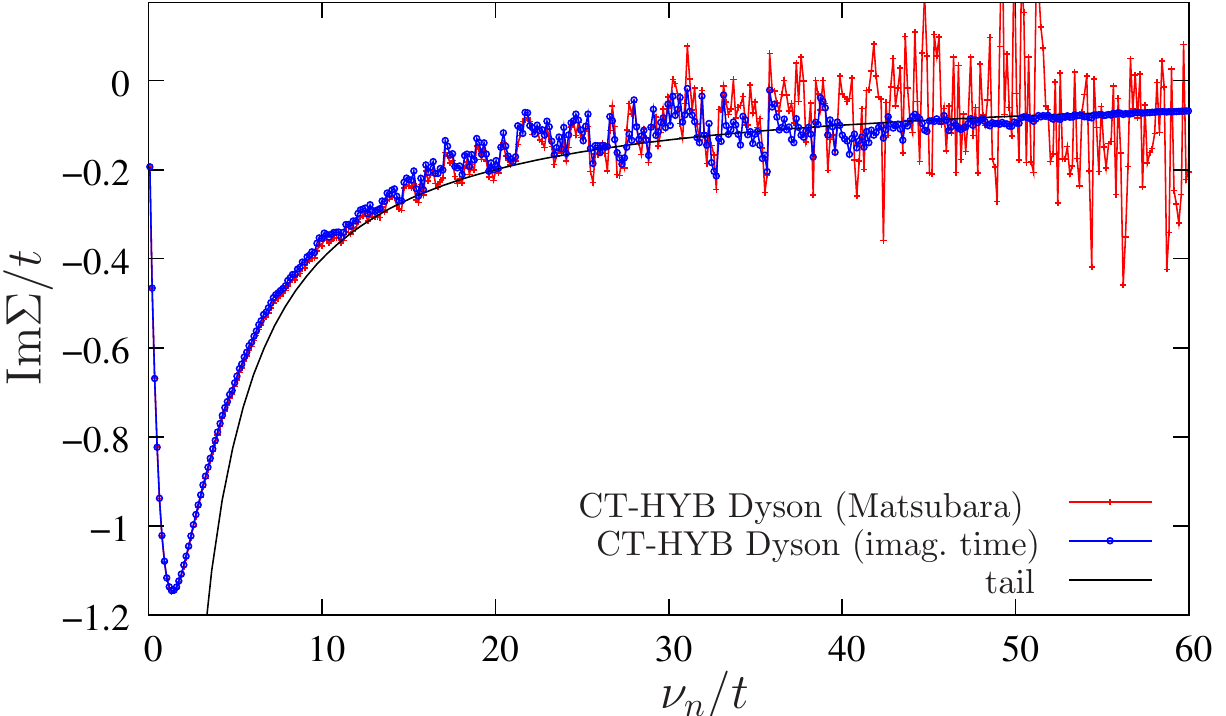} 
\caption{\label{fig:sigma_before} (Color online) Self-energy as a function of Matsubara frequencies for the half-filled Hubbard model on the Bethe lattice (bandwidth $4t$) as calculated from Dyson's equation to illustrate the noise problem. The parameters are the same as in Ref.~\onlinecite{Gull07}, $U/t=4$ and $T/t=1/45$. The results were obtained by measuring the Green's function directly on Matsubara frequencies and in imaginary time using 1500 bins and subsequent Fourier transform for the latter. The high-frequency tail $\lim_{\nu\to\infty}\Sigma(\iom)= U^{2} \av{n}(1-\av{n})/(\iom)$ is shown for comparison. 
}
\end{figure}

A contribution to the Green's function of a particular Monte Carlo configuration is obtained by cutting all hybridization lines connected to a given pair of a creation and an annihilation operator, leaving a configuration with two unconnected operators. This corresponds to removing row $i$ and column $j$ from the matrix of hybridization functions $\hat{\Delta}^{\mathcal{C}}$. Denoting the resulting matrix as $\hat{\Delta}^{\mathcal{C}}_{ij}$, the contribution of the particular configuration is essentially the ratio between the matrices after and before removing the hybridization functions, i.e. $(-1)^{i+j}\det \hat{\Delta}^{\mathcal{C}}_{ij}/\det\hat{\Delta}^{\mathcal{C}}$. This ratio is precisely the $j,i$-element of the inverse of the matrix of hybridization functions, denoted by $M^{\mathcal{C}}_{ji}$. Hence the Green's function defined on the interval from 0 to $\beta$ is measured according to\cite{Werner06,WernerPRBLong}
\begin{align}
&G_{ab}(\tau-\tau') = \nonumber\\ 
&-\frac{1}{\beta}\left\langle \sum_{\alpha\beta=1}^{k^{\mathcal{C}}} M^{\mathcal{C}}_{\beta\alpha}\delta^{-}(\tau-\tau',\tau_{\alpha}^{e}-\tau_{\beta}^{s})\delta_{a,\alpha}\delta_{b,\beta}\right\rangle_{\text{MC}} ,
\end{align}
where 
\begin{equation}
\delta^{-}(\tau,\tau') \Let \sgn(\tau')\delta(\tau-\tau' -\theta(-\tau')\beta).
\label{eqn:deltam_def}
\end{equation}
The only difference in the measurement for the function $F_{ab}^{j}(\tau-\tau')$ is the presence of the additional density operator. Hence the measurement formula reads
\begin{align}
&F_{ab}^{j}(\tau-\tau') =\nonumber\\ 
&-\frac{1}{\beta}\left\langle \sum_{\alpha\beta=1}^{k^{\mathcal{C}}} M_{\beta\alpha}^{\mathcal{C}}\delta^{-}(\tau-\tau',\tau_{\alpha}^{e}-\tau_{\beta}^{s}) n_{j}(\tau_{\beta}^{s})\delta_{a,\alpha}\delta_{b,\beta}\right\rangle_{\text{MC}} .
\end{align}
In the segment picture, the matrix element $n_{j}(\tau_{\beta}^{s})$ (one or zero) of this operator is simply determined by examining whether or not a segment is present in the timeline for flavor $j$ at time $\tau_{\beta}^{s}$. Hence this quantity can be measured at essentially no additional computational cost.
Note that this function according to \eqref{eqn:sigmafinal} only contributes if $j$ is different from the index $b$ (and $\beta$) and therefore $n_{j}$ is never evaluated  at the position of the creator of flavor $b$ at time $\tau_{\beta}^{s}$.

\begin{figure}[t]
\includegraphics[scale=0.675,angle=0]{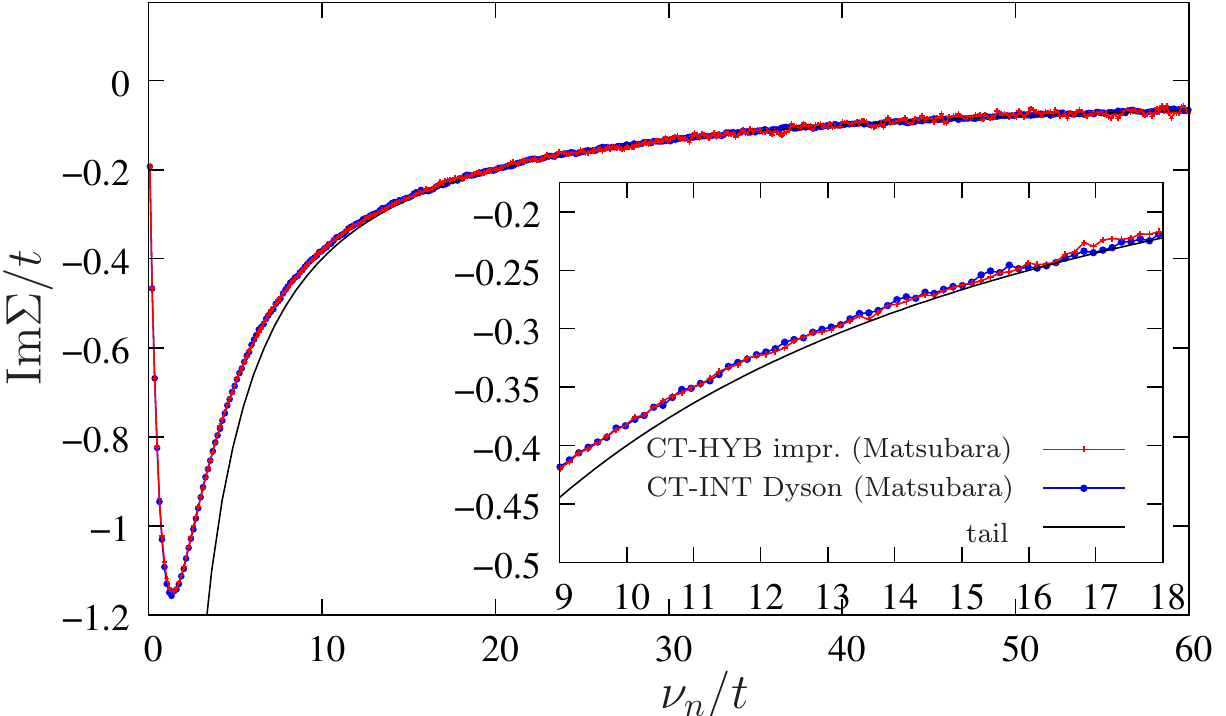} 
\caption{\label{fig:sigma_improved} (Color online) CT-HYB data for the self-energy computed from Eq.~\eqref{eqn:sigmafinal} for the same model and parameters and measured within the same simulation as the results in Fig.~\ref{fig:sigma_before}. The Green's function and the two-particle correlator have been measured on the Matsubara axis. The line with closed symbols (blue) shows the result from an independent DMFT calculation using the interaction expansion algorithm (CT-INT), where the self energy has been obtained via Dyson's equation with the Green's function measured on the Matsubara axis. The noise problem at intermediate and high frequencies does not exist in the CT-INT and is resolved in CT-HYB using the improved estimator. The inset shows a blowup of the intermediate frequency range.}
\end{figure}

It is convenient to accumulate 
\begin{equation}
(G \Sigma)_{ab}(\tau-\tau') = (1/2) \sum_{j} (U_{jb}+U_{bj}) F^{j}_{ab}(\tau-\tau')
\label{eqn:sigmag}
\end{equation}
directly instead of the individual quantities $F^{j}_{ab}(\tau-\tau')$.
This is then analogous to the measurement of $G$ times $\Sigma$ in the CT-AUX algorithm.\cite{CTAUX} Note that these correlators can be measured in any basis by appropriately transforming the measurement rules, e.g. by taking the Fourier transform to measure directly the Matsubara coefficients.

If the interaction is of non-density-density type, the segment picture has to be abandoned. In this case, the equation of motion generates additional terms, which require the measurement of correlation functions of the form $\av{\T c_{a}(\tau)c_{b}^{\dagger}(\tau') c_{j}^{\dagger}(\tau') c_{k}(\tau')}$ (see appendix \ref{eqn:app:gencorr}). In addition to computing the ratio of determinants, the measurement of such functions requires the evaluation of the ratio of traces over the operators in the particular configuration with and without the operators $c_j^{\dagger}(\tau'), c_{k}(\tau')$ inserted at the corresponding time.

\begin{figure}[t]
\includegraphics[scale=0.675,angle=0]{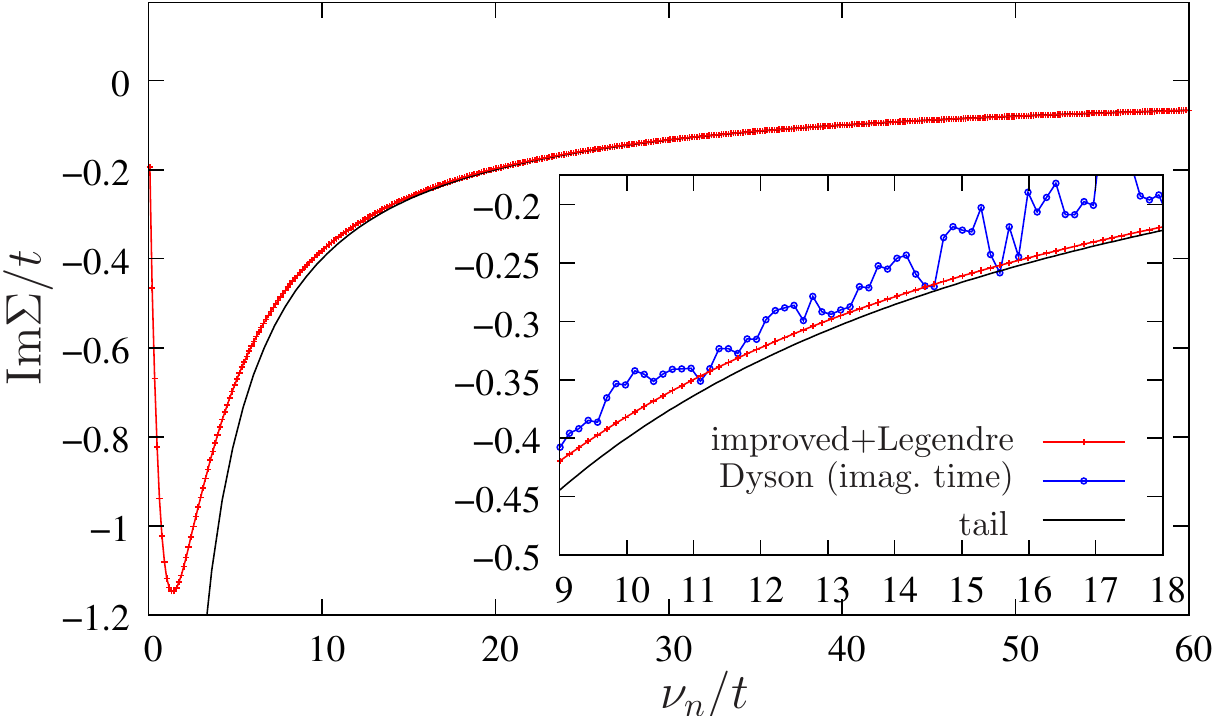} 
\caption{\label{fig:sigma_legendre} (Color online) Same as in Fig. \ref{fig:sigma_improved}, albeit with the Green's function and the two-particle correlator measured in the Legendre polynomial basis and transformed back to Matsubara frequencies. This allows to efficiently filter the Monte Carlo noise. 
The inset compares the result at intermediate frequencies to that of Fig. \ref{fig:sigma_before} obtained from the Fourier transformed imaginary time measurement (data are measured within the same simulation).}
\end{figure}

The major benefit of rewriting the self-energy in the form \eqref{eqn:sigmafinal} instead of computing it from Dyson's equation
\begin{align}
\Sigma(\iom) = G_{0}(\iom)^{-1} - G(\iom)^{-1}
\label{eqn:dyson}
\end{align}
is that it is expressed in terms of a \emph{ratio} of two measured quantities instead of a \emph{difference} of two functions. Forming the difference between an exactly known and an approximate quantity is susceptible to numerical errors, since, as already noted in Ref. \onlinecite{NRG}, for a ratio only the relative error propagates, while in a difference, the absolute error (here the absolute error of $G^{-1}$) propagates. 
In DMFT, $G_{0}$ is computed from the self-energy of the previous iteration and is not known exactly. Error cancellation however cannot be expected since $G_0$ and $G$ are computed in two independent Monte Carlo runs.
Since $G$ decays as $1/(\iom)$ it is clear that computing the self-energy according to Eq.~\eqref{eqn:dyson} will lead to large errors in particular at intermediate to large frequencies.

In the following we show results illustrating the advantages of the improved measurement for the CT-HYB. We concentrate on the DMFT solution of the Hubbard model on the Bethe lattice with bandwidth $4t$. The parameters are the same as in Ref.~\onlinecite{Gull07}. In order to ensure comparability, we have performed the various measurements within the same simulation, i.e. all quantities have been averaged over the identical sequence of Monte Carlo configurations (including the results of Fig. \ref{fig:sigma_before}).

Figure~\ref{fig:sigma_before} shows the self-energy determined in the standard way, i.e. by measuring the Green's function and calculating the self-energy using Dyson's equation. The Monte Carlo noise is clearly visible for intermediate to large frequencies, as anticipated. It is important to note that there is also considerable noise in the region where the self-energy has clearly not yet reached its asymptotic behavior (cf. also inset of Fig. \ref{fig:sigma_legendre} below). 
While the results are similar for the frequency and imaginary time measurement at small to intermediate frequencies, the Monte Carlo error steadily grows with the frequency for the measurement in Matsubara frequencies.
The noise sets in much below the Nyquist frequency (of approximately $\nu_{n}/t\sim 50$). In the imaginary-time measurement the noise is suppressed above the Nyquist frequency and the result smoothly approaches the tail, which is enforced in the calculation of the Fourier transform.

\begin{figure}[t]
\includegraphics[scale=0.675,angle=0]{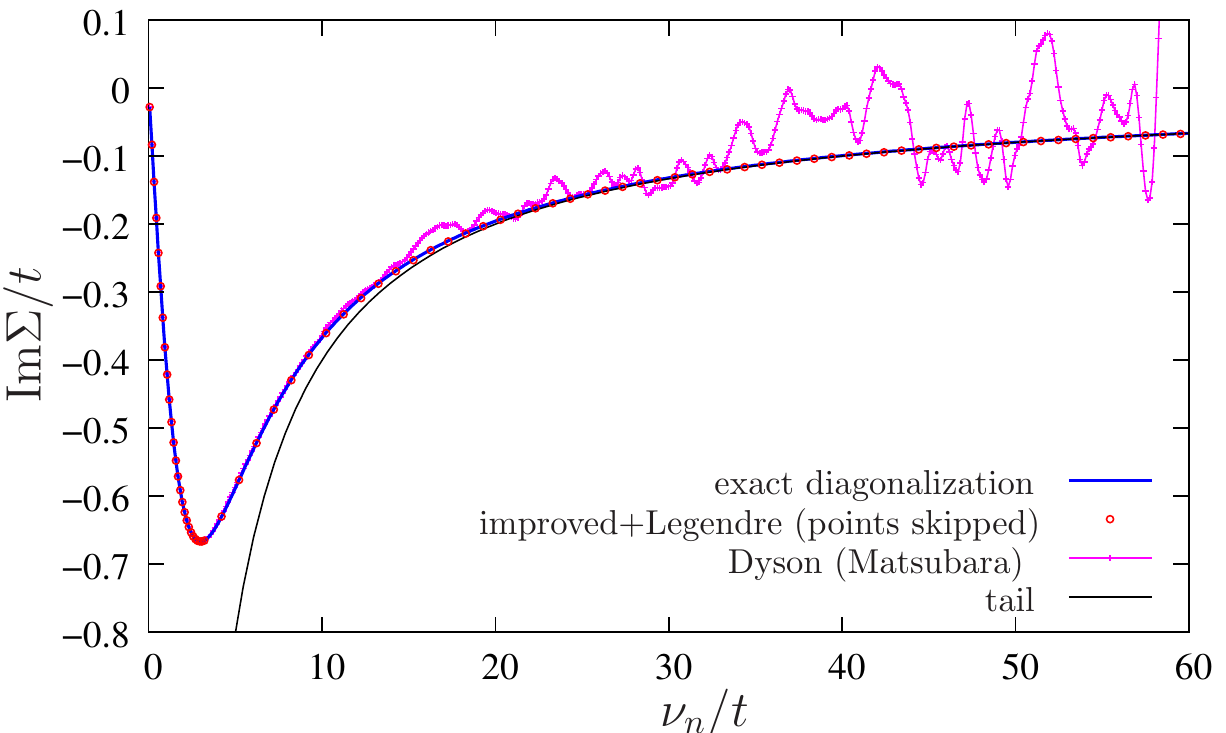} 
\caption{\label{fig:sigma_ed} (Color online) Comparison of Monte Carlo data for the imaginary part of the self-energy (obtained with the same method as in Fig.~\ref{fig:sigma_legendre}) with the exact result for a correlated site coupled to a single bath level. Some points of the Monte Carlo result are skipped to improve visualization. The result computed from Dyson's equation (with the Green's function measured on the Matsubara axis within the same simulation) illustrates the substantial improvement in accuracy.
}
\end{figure}

Figure \ref{fig:sigma_improved} clearly shows the advantage of the improved measurement: The error is considerably reduced over the entire frequency range. The number of measurements required for a given accuracy at intermediate and large frequencies is hence greatly reduced.
In the same figure we compare the improved measurement of CT-HYB with a result obtained from a comparable run of the CT-INT using Dyson's equation, with the Green's function also measured on the Matsubara axis. 
We note that this is meant as an illustration of the convergence properties at high frequency. Here we do not attempt the intricate task of separating the various effects that influence the performance and accuracy of these complementary algorithms, which furthermore scale very differently with the number of flavors.\cite{CTQMCRMP} Comparable refers to the fact that for both runs we have used similar simulation times that yield converged results in practice. A detailed performance comparison of the algorithms was presented in Ref.~\onlinecite{Gull07}.

The result illustrates that the noise problem does not exist in the CT-INT algorithm. 
The reason is that in CT-INT, the Green's function is measured as an $O((1/\iom)^2)$ correction to the bare Green's function $G_{0}$. As a consequence, the evaluation of the self-energy is considerably more stable (see below). We find no advantage of calculating the self-energy according to Eq.~\eqref{eqn:sigmafinal} in CT-INT, at least for the model considered here.

We note that in principle one may evaluate the self-energy without explicitly measuring the Green's function $G_{ab}(\iom)$.
Combining Eqs. \eqref{eqn:dyson1} and \eqref{eqn:sigmafinal}, one sees that the Green's function may be written in the form
\begin{equation}
G_{ab}(\iom) = \sum_{a'} A_{aa'} G^{0}_{a'b}(\iom),
\label{eqn:galt}
\end{equation}
where the matrix $A_{aa'}$ is defined as
\begin{align}
A_{aa'} = \delta_{aa'} + \frac{1}{2}\sum_{j}(U_{ja'}+U_{a'j})F_{aa'}^{j} .
\label{eqn:Aaa}
\end{align}
The self-energy is obtained by multiplying \eqref{eqn:sigmag} by the inverse of \eqref{eqn:galt}. We find that this procedure yields results of similar but somewhat worse quality than the ones shown in Fig. \ref{fig:sigma_improved}, so that it does not reduce the effective computation time for given accuracy.

In Fig.~\ref{fig:sigma_legendre} we show results obtained by combining the improved estimator with an efficient method to suppress the Monte Carlo noise. The latter is based on a representation in terms of Legendre orthogonal polynomials. The transformation to the Legendre basis is exact, as in the Fourier case. Both $G(\iom)$ and $F(\iom)$ have been measured directly in the Legendre basis and transformed back exactly to Matsubara frequencies after the simulation.
Appropriately choosing the cutoff in this basis allows to eliminate the Monte Carlo noise without loosing physical information (for details see Ref.~\onlinecite{legendre}).
This method yields very accurate, noise-free results over the entire frequency range and captures the high-frequency tail correctly.
We expect that Monte Carlo data measured in this way will allow a more stable analytical continuation.

In order to show that the proposed method not only yields smooth, but indeed correct results, we compare the Monte Carlo data to an exact result.
A suitable test case is a correlated impurity site coupled to a single bath level at $\epsilon=0$ for which the Hamiltonian (\ref{eqn:hamiltonian}) reduces to 
\begin{equation} 
H = t[c^\dagger f + f^\dagger c] + U n_\uparrow n_\downarrow 
\end{equation} 
and which is trivially solved by exact diagonalization. The hybridization function for this case is $\Delta(\tau)=t^2/2=\text{const.}$. 
Although the hybridization function has no structure in this case, this is not a trivial problem for the Monte Carlo approach, since all diagrams in principle have to be sampled to obtain the exact solution. The result at half-filling is shown in Fig.~\ref{fig:sigma_ed} for $U/t=4$ and temperature $T/t=1/50$. To improve the visualization, we have skipped some points from the calculated result. The curves lie on top of each other. That this is indeed not trivial can be seen from the result computed from Dyson's equation using a Green's function measurement on the Matsubara axis, which has again been accumulated within the same simulation.

\subsection{Vertex function}

\begin{figure}[b]
\includegraphics[scale=0.475,angle=0]{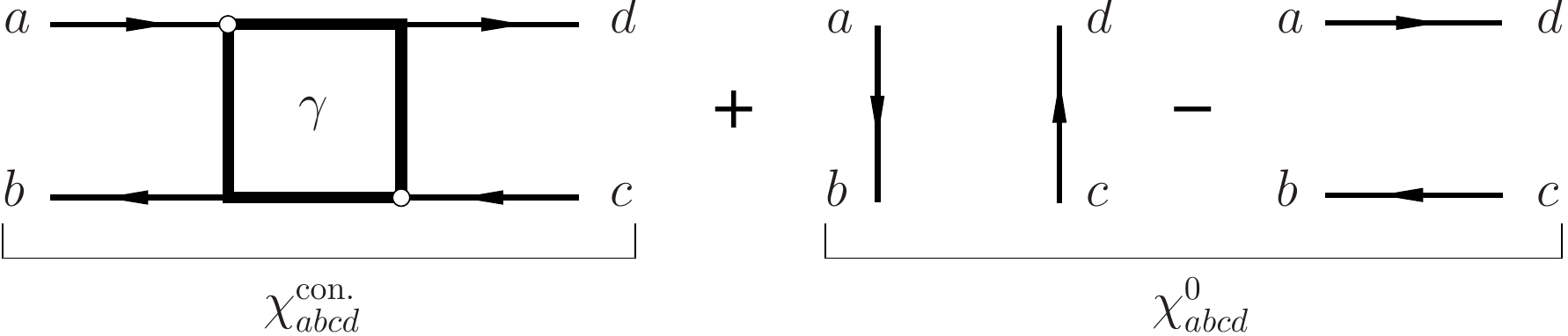} 
\caption{\label{fig:2pgf_def} Diagrammatical representation of the two-particle Green's function $\chi_{abcd}$ in terms of its connected part $\chi^{\text{con.}}_{abcd}$ and its disconnected part $\chi^{0}_{abcd}$ and definition of the vertex function $\gamma$, Eqs. \eqref{eqn:chi0}-\eqref{eqn:gamma}. The lines with arrows denote fully dressed propagators.
}
\end{figure}

\begin{figure}[t]
\includegraphics[scale=0.675,angle=0]{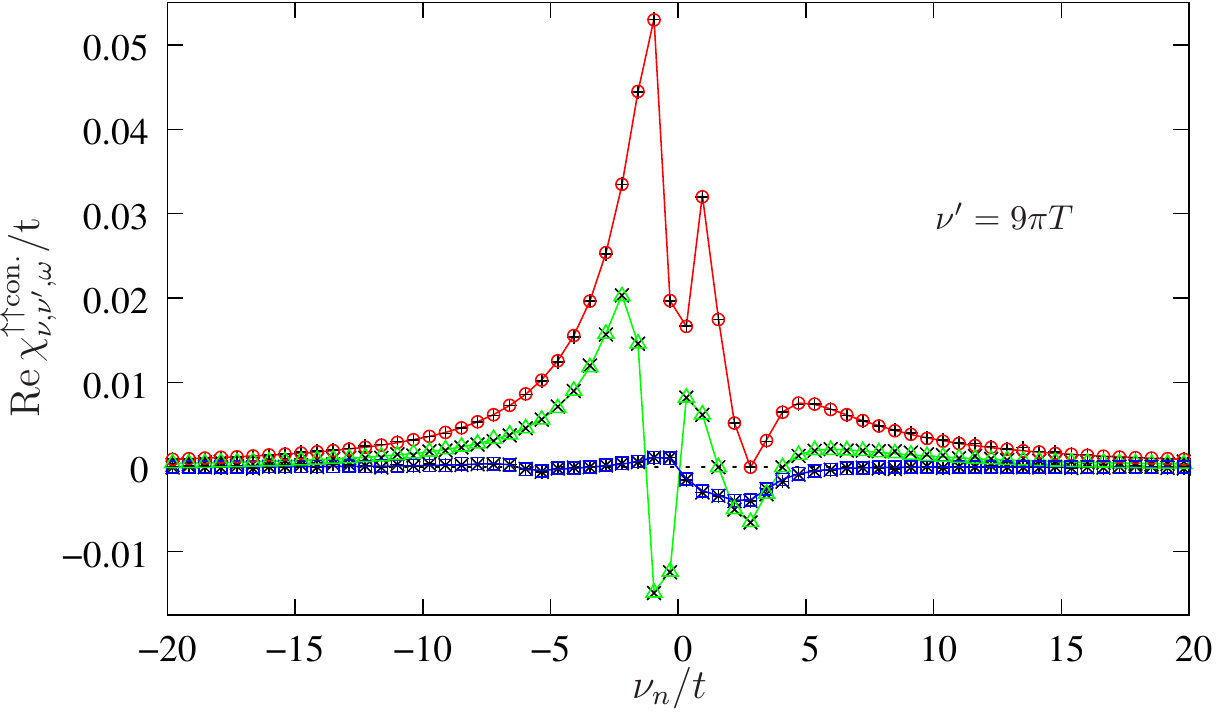} 
\caption{\label{fig:g2_irr_b10} (Color online) Real part of the spin-up-up component of the connected part of the two-particle Green's function for a correlated impurity coupled to a single bath-level for $T/t=1/10$ for fixed $\nu'$ and bosonic frequencies $\omega=0$ (circles), $4\pi T$ (triangles) and $20\pi T$ (squares). The parameters are otherwise the same as in Fig. \ref{fig:sigma_ed}. Solid lines show the exact diagonalization data and open symbols the results from the improved Monte Carlo measurement (measured on the Matsubara axis).
The usual Matsubara axis measurement based on Eqs.~\eqref{eqn:chi}-\eqref{eqn:gamma} is shown by crosses. The two results are hardly distinguishable on this scale.
}
\end{figure}

To set the stage for the discussion of the vertex function, we define the two-particle Green's function as
\begin{align}
\chi_{abcd}(\tau_a,\tau_b,\tau_c,\tau_d) \Let  \av{\T c_a(\tau_a) c_b^\dagger(\tau_b) c_c(\tau_c) c_d^\dagger(\tau_d)}.
\label{eqn:chi}
\end{align}
Its Fourier transform is $\chi_{abcd}(\iom_{a},\iom_b,\iom_c,\iom_d)$. With the disconnected part of the two-particle Green's function,
\begin{align}
\chi_{abcd}^{0}(\iom_{a},\iom_b,\iom_c,\iom_d) \Let &G_{ab}(\iom_{a})\delta_{\nu_{a},\nu_{b}}G_{cd}(\iom_{c})\delta_{\nu_{c},\nu_{d}} \nonumber\\&-\!G_{ad}(\iom_{a})\delta_{\nu_{a},\nu_{d}}G_{cb}(\iom_{c})\delta_{\nu_{c},\nu_{b}} 
\label{eqn:chi0}
\end{align}
we obtain the connected part (cf. Fig. \ref{fig:2pgf_def}):
\begin{align}
\chi^{\text{con.}}_{abcd} (\iom_a,\iom_b,\iom_c,\iom_d) \Let &\chi_{abcd}(\iom_a,\iom_b,\iom_c,\iom_d) \nonumber\\ 
& - \chi_{abcd}^{0}(\iom_a,\iom_b,\iom_c,\iom_d),
\label{eqn:chiconn}
\end{align}
in terms of which the two-particle vertex function is defined as
\begin{align}
&\gamma_{abcd}(\iom_a,\iom_b,\iom_c,\iom_d) \Let
G^{-1}_{aa'}(\iom_{a})G^{-1}_{cc'}(\iom_{c})\times\nonumber\\
&\quad \times\chi^{\text{con.}}_{a'b'c'd'} (\iom_a,\iom_b,\iom_c,\iom_d) 
G^{-1}_{b'b}(\iom_{b})G^{-1}_{d'd}(\iom_{d}).
\label{eqn:gamma}
\end{align}

\begin{figure}[b]
\includegraphics[scale=0.675,angle=0]{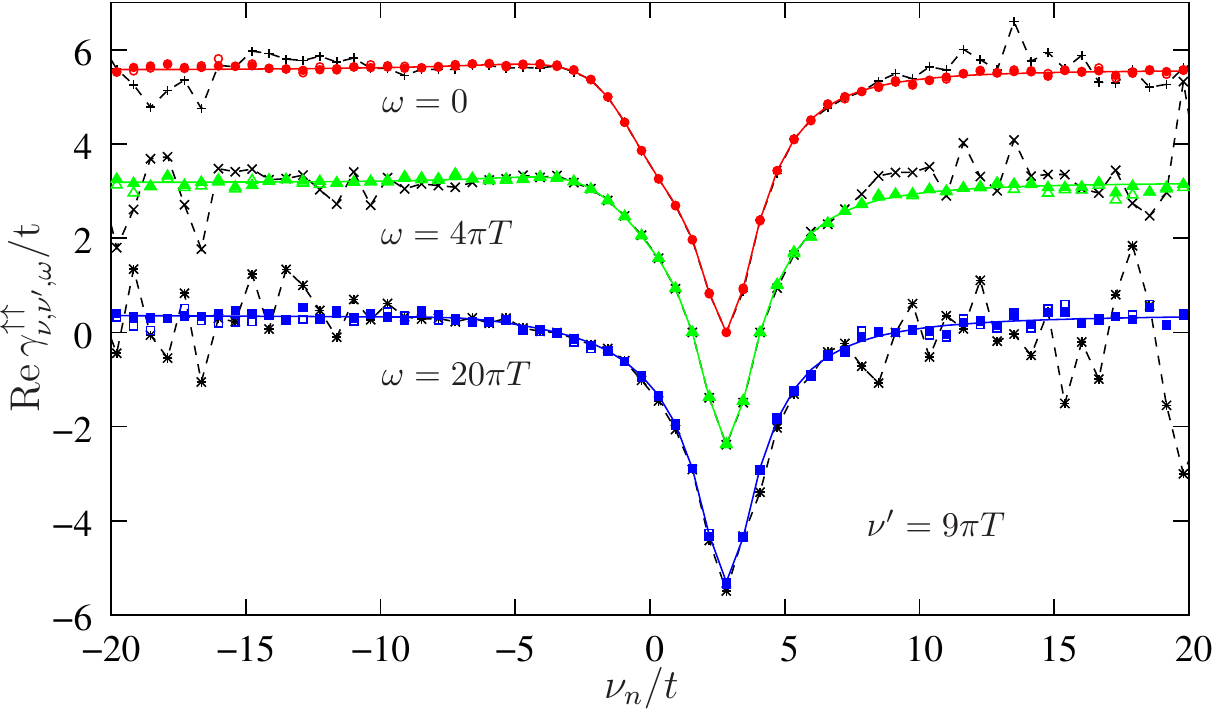} 
\caption{\label{fig:gamma_b10} (Color online) Real part of the spin up-up component of the reducible vertex function $\gamma_{\nu\nu'\omega}$ computed from the 
connected part shown in Fig. \ref{fig:g2_irr_b10}.
The standard computation (Eqs. \eqref{eqn:chi}-\eqref{eqn:gamma}, crosses connected by dashed lines) is clearly numerically unstable for large values of $\nu_{n}$, while the improved estimators yield the proper asymptotic behavior.
The improved estimator computed from Eqs.~\eqref{eqn:gamma}, \eqref{eqn:chiconnfinal} (closed symbols) and from Eqs.~\eqref{eqn:gamma}, \eqref{eqn:chiconnalt} (open symbols) yield similar results.
They also yield more accurate results even \emph{before} the vertex approaches its asymptotic behavior and for small frequencies $\nu_{n}$ (clearly visible for $\omega=20\pi T$).
}
\end{figure}

The connected part of the two-particle Green's function is the difference between a two-particle quantity and a product of two single-particle Green's functions. The calculation of this difference is susceptible to numerical errors. Since single- and two-particle quantities have vastly different statistical errors, one cannot expect these to cancel. The error in the vertex function itself is further amplified by multiplying the connected part with inverse Green's functions. This leads to deviations in particular when $G$ is small, i.e. for large frequencies or in an insulator, where $G(\iom)$ extrapolates to zero for $\iom\rightarrow 0$.

Hence we seek to express the connected part in a similar fashion as was done for the self-energy using the equation of motion.
As shown in appendix \ref{app:vertex}, the connected part of the two-particle Green's function can be written in the alternative form
\begin{align}
&\chi^{\text{con.}}_{abcd} (\iom_a,\iom_b,\iom_c,\iom_d) =\nonumber\\
&-\frac{1}{2}\sum_{ij}(U_{ji}+U_{ij}) \Big[F^j_{ai}(\iom_a)\, \chi_{ibcd}(\iom_a,\iom_b,\iom_c,\iom_d)\nonumber\\
&\qquad\qquad\qquad\qquad\ -G_{ai}(\iom_a) H^j_{ibcd}(\iom_a,\iom_b,\iom_c,\iom_d)\Big].
\label{eqn:chiconnfinal}
\end{align}
A priori it is not clear that computing the connected part according to this expression has any advantage compared to Eq.~\eqref{eqn:chiconn}, since it still has the form of a difference of products of correlation functions. However we find that it indeed yields more accurate results.

In addition to the single- and two-particle Green's function this expression involves the  correlation functions $F_{ab}^{j}(\iom)$ and the Fourier transform of
\begin{align}
H_{abcd}^j(\tau_a,\tau_b,\tau_c,\tau_d) &\Let \av{\T n_j(\tau_a) c_{a}(\tau_a) c_b^\dagger(\tau_b) c_c(\tau_c) c_d^\dagger(\tau_d)}.
\end{align}
In order to see how this function is measured, recall that in CT-HYB the two-particle Green's function with its arguments varying in the interval from 0 to $\beta$ is measured according to\cite{CTQMCRMP}
\begin{align}
&\chi_{abcd}(\tau_{ab},\tau_{cd},\tau_{ad}) = \nonumber\\ 
&\frac{1}{\beta}\Bigg\langle\! \sum_{\alpha\beta\gamma\delta=1}^{k^{\mathcal{C}}} (M^{\mathcal{C}}_{\beta\alpha}M^{\mathcal{C}}_{\delta\gamma}- M^{\mathcal{C}}_{\delta\alpha}M^{\mathcal{C}}_{\beta\gamma})\delta^{-}(\tau_{ab},\tau_{\alpha}^{e}-\tau_{\beta}^{s})\times\nonumber\\
&\ \times\delta^{-}(\tau_{cd},\tau_{\gamma}^{e}-\tau_{\delta}^{s})\delta^{+}(\tau_{ad},\tau_{\alpha}^{e}-\tau_{\delta}^{s})\delta_{a,\alpha}\delta_{b,\beta}\delta_{c,\gamma}\delta_{d,\delta}
\Bigg\rangle_{\text{MC}} ,
\label{eqn:measG2}
\end{align}
where $\delta^{-}$ is defined as before \eqref{eqn:deltam_def} and $\delta^{+}(\tau,\tau') \Let \delta(\tau-\tau' -\theta(-\tau')\beta)$.
Due to time translation invariance, the two-particle Green's function needs to be measured as a function of three independent time differences only. These have been chosen such that $\chi$ is antiperiodic in $\tau_{ab}$ and $\tau_{cd}$, while it is periodic in $\tau_{ad}$. When taking the Fourier transform, the time differences $\tau_{ab}$ and $\tau_{cd}$ are associated with fermionic Matsubara frequencies $\nu$, $\nu'$, while $\tau_{ad}$ is associated with a bosonic frequency $\omega$. Note that the relation to the representation with four frequencies is such that
\begin{align}
\chi_{abcd}(\nu,\nu',\omega) 
\equiv \chi_{abcd}(\iom+i\omega,\iom,\iom',\iom'+i\omega).
\end{align}
The measurement formula for the correlator $H^{j}_{abcd}$ reads
\begin{align}
&H^{j}_{abcd}(\tau_{ab},\tau_{cd},\tau_{ad}) = \nonumber\\ 
&\frac{1}{\beta}\Bigg\langle\! \sum_{\alpha\beta\gamma\delta=1}^{k^{\mathcal{C}}} (M^{\mathcal{C}}_{\beta\alpha}M^{\mathcal{C}}_{\delta\gamma}- M^{\mathcal{C}}_{\delta\alpha}M^{\mathcal{C}}_{\beta\gamma})n_{j}(\tau_{\alpha}^{e})\delta^{-}(\tau_{ab},\tau_{\alpha}^{e}-\tau_{\beta}^{s})\times\nonumber\\
&\ \times\delta^{-}(\tau_{cd},\tau_{\gamma}^{e}-\tau_{\delta}^{s})\delta^{+}(\tau_{ad},\tau_{\alpha}^{e}-\tau_{\delta}^{s})\delta_{a,\alpha}\delta_{b,\beta}\delta_{c,\gamma}\delta_{d,\delta}
\Bigg\rangle_{\text{MC}} ,
\label{eqn:measH}
\end{align}
in analogy to $F_{ab}^{j}$. The correlator $H^{j}_{abcd}$ can thus be measured at essentially no additional computational cost together with the two-particle Green's function.

\begin{figure}[t]
\includegraphics[scale=0.675,angle=0]{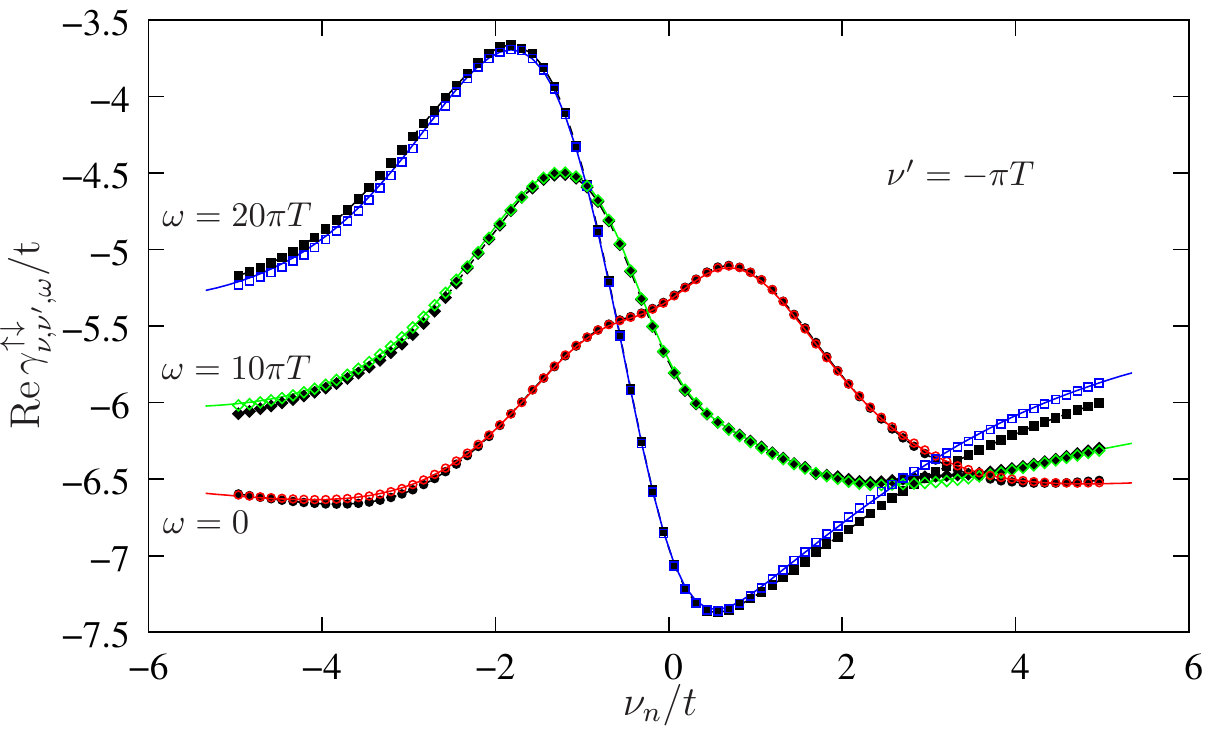} 
\caption{\label{fig:vertex_legendre} (Color online) Spin-up-down component of the real part of the reducible vertex function for a correlated impurity coupled to a single bath-level. The parameters are the same as in Fig. \ref{fig:sigma_ed}. Lines show the exact diagonalization result and open symbols show the improved Monte Carlo measurement. Black symbols correspond to the usual Matsubara axis measurement based on Eqs.~\eqref{eqn:chi}-\eqref{eqn:gamma}
(note that the measurement in the Legendre basis leads to systematic deviations rather than the irregular noise observed in the Matsubara measurement).
}
\end{figure}

Single-particle quantities can be measured directly in imaginary time with essentially arbitrarily fine resolution, since the number of imaginary time bins does not influence the performance of the algorithm. For two-particle quantities, which depend on three independent time differences, this is not the case, due to memory restrictions. Hence we measure these quantities directly in the Matsubara or the Legendre basis. 
Instead of measuring the correlator $H^{j}(\nu,\nu',\omega)$  for all flavors $j$, we accumulate $(1/2) \sum_{j} (U_{ja}+U_{aj}) H^{j}_{abcd}(\nu,\nu',\omega)$ in order to save memory. However, this still requires one to measure, in addition to the two-particle Green's function, an object  which is of the same size. 
In order to avoid this, one can replace $\chi_{abcd}$ on the right hand side of Eq.~\eqref{eqn:chiconnfinal} by the sum of its connected and disconnected parts, which leads to
\begin{align}
&\chi^{\text{con.}}_{abcd} (\iom_a,\iom_b,\iom_c,\iom_d) =
\sum_{a'}A^{-1}_{aa'}\Bigg\{\nonumber\\ 
&-\frac{1}{2}\sum_{ij}(U_{ji}+U_{ij})
\Big[F^j_{a'i}(\iom_a)\, \chi^{0}_{ibcd}(\iom_a,\iom_b,\iom_c,\iom_d)\nonumber\\
&\qquad\qquad\qquad\qquad -G_{a'i}(\iom_a) H^j_{ibcd}(\iom_a,\iom_b,\iom_c,\iom_d)\Big]\Bigg\},
\label{eqn:chiconnalt}
\end{align}
where $A_{aa'}$ is defined as in \eqref{eqn:Aaa}.
Note that here the disconnected part instead of the full two-particle Green's function appears on the right hand side.
In the following we assess the quality of the results obtained from the different measurement formulas for the vertex function.

As a test, we again consider a correlated impurity hybridized to a single bath level. This allows us to compare Monte Carlo data to exact results. Figure~\ref{fig:g2_irr_b10} shows the real part of the spin-up-up component of the connected part of the two-particle Green's function,
$\text{Re}\chi^{\uparrow\uparrow\text{con.}}_{\nu,\nu',\omega}$, as a function of the fermionic Matsubara frequency $\nu_n$ for fixed $\nu'=9\pi T$ ($T/t=10$, $U/t=4$). Circles, squares and triangles correspond to the bosonic frequencies $\omega=0$, $4\pi T$, and $20\pi T$, respectively. The improved Monte Carlo measurements (open and closed symbols), as well as the usual Matsubara-axis measurements (crosses) agree very well with the exact diagonalization result (lines). The problems caused by the stochastic noise are not evident at the level of the two-particle Green's function, but only at the level of the vertex. 

The real part of the spin-up-up component of the reducible vertex is shown in Fig.~\ref{fig:gamma_b10}. Here, due to the multiplication with inverse Green's functions, the noise in the data obtained from the standard Matsubara measurement grows considerably at large frequencies, while the improved measurement reproduces the correct high-frequency behavior. Both improved estimators, Eqs.~\eqref{eqn:gamma}, \eqref{eqn:chiconnfinal} and Eqs.~\eqref{eqn:gamma}, \eqref{eqn:chiconnalt}, yield results of comparable accuracy. In the case of larger bosonic frequencies, where the connected part becomes small, the improved accuracy of the new measurement procedure is evident even at the lowest $\nu_n$.

We found that using the vertex function obtained via the improved estimators enhances the stability of dual fermion calculations. This is particularly the case for calculations in the insulating phase, 
where the improved estimators appear to be significantly more accurate and where 
it is otherwise difficult to obtain sufficient statistics due to the low perturbation order.

Figure~\ref{fig:vertex_legendre} shows the real part of the spin-up-down component of the reducible vertex at low temperature and for the same parameters as in Fig.~\ref{fig:sigma_ed} ($T/t=1/45$, $U/t=4$), measured in the Legendre basis. Again, one sees that at large bosonic $\omega_m$, deviations between the exact result (lines) and the standard Matsubara measurement (black symbols) appear already at small $\nu_n$, while the improved estimators (symbols) yield more accurate data. 

\section{Application}

\subsection{Two-orbital model}

As an application of the methods described above, we calculate the self-energy and vertex function for a two-orbital model within the single-site DMFT. We consider the Hubbard model on the Bethe lattice with semi-elliptical density of states with bandwidth $4t$. First, we re-examine the spin-freezing transition reported in Ref.~\onlinecite{spinfreezing} for a three-orbital model, and show that qualitatively the same physics is found in the two-orbital model, with the interaction restricted to density-density terms.
We will work with the Hamiltonian \eqref{eqn:hamiltonian}, with the interaction part  explicitly given by
\begin{align}
H_{U} =&\ U\!\!\sum_{\alpha=1,2} n_{\alpha\uparrow}n_{\alpha\downarrow} + U'\sum_{\sigma}n_{1,\sigma}n_{2,-\sigma} \nonumber\\
& + (U'-J) \sum_{\sigma} n_{1,\sigma}n_{2,\sigma}, 
\end{align}
where $\alpha$ is the orbital index, $\sigma$ denotes spin, $U$ and $U'$ are the intra- and inter-orbital Coulomb interaction parameters, $J$ is the Hund's rule coupling coefficient and $U'=U-2J$.

\subsection{Spin-freezing transition}

\begin{figure}[t]
\includegraphics[scale=0.675,angle=0]{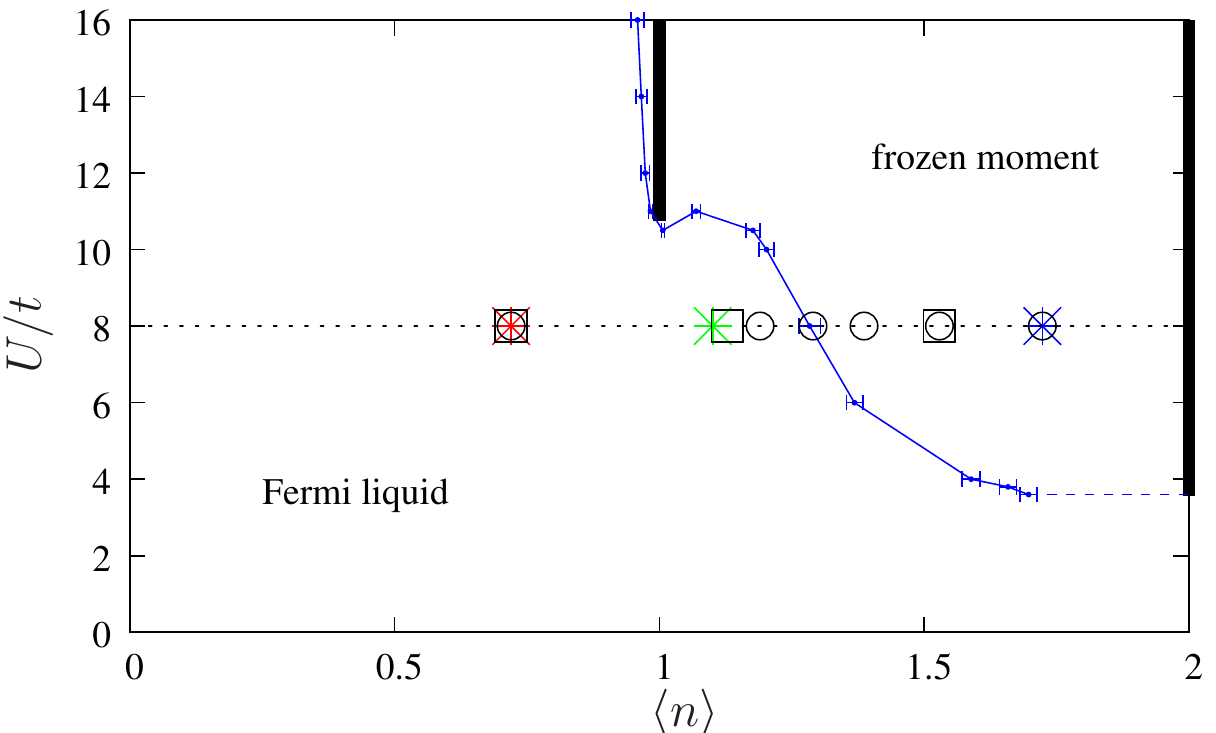} 
\caption{\label{fig:phasediagram_sf} (Color online) Phasediagram in the space of density $\av{n}$ and interaction $U$ of the two-orbital Hubbard model on the Bethe lattice  with $J/U=1/6$ and $T/t=0.02$. The thick black lines mark regions of Mott insulating behavior. Circles, squares and stars, respectively, mark the parameters for which the calculations in Figs. \ref{fig:sigma_sf}, \ref{fig:sigmaE} and \ref{fig:vertex_sf} have been performed.
}
\end{figure}

\begin{figure}[t]
\includegraphics[scale=0.675,angle=0]{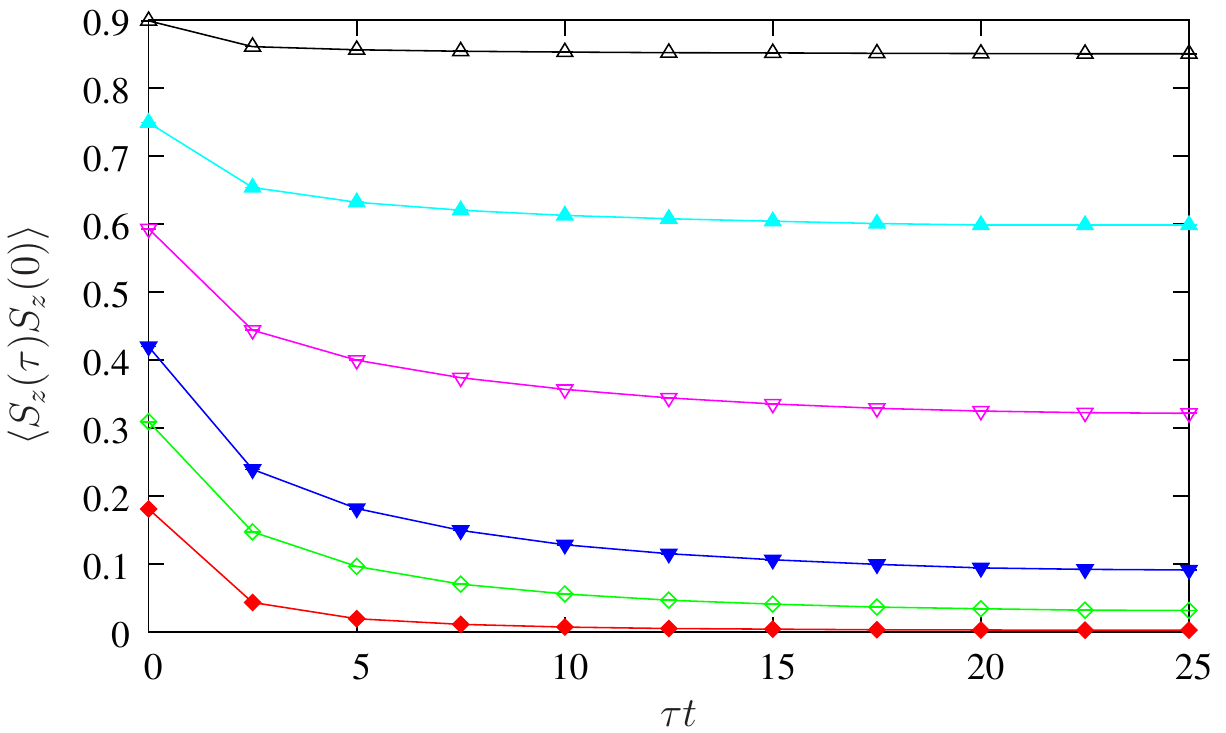} 
\caption{\label{fig:szsz_tau} (Color online) Imaginary-time dependence of the spin-spin correlation function for $U/t=8$ and densities $\av{n}=1.90,1.72,1.53,1.29,1.10$ and $0.72$ (from top to bottom). For small densities, in the Fermi liquid regime, the spin-spin correlation at $\tau=1/(2T)$ is proportional to $T$ and hence approaches zero as $T\to 0$, while for high densities the spin correlation function approaches a constant, indicating the presence of frozen moments.
}
\end{figure}

\begin{figure}[t]
\includegraphics[scale=0.675,angle=0]{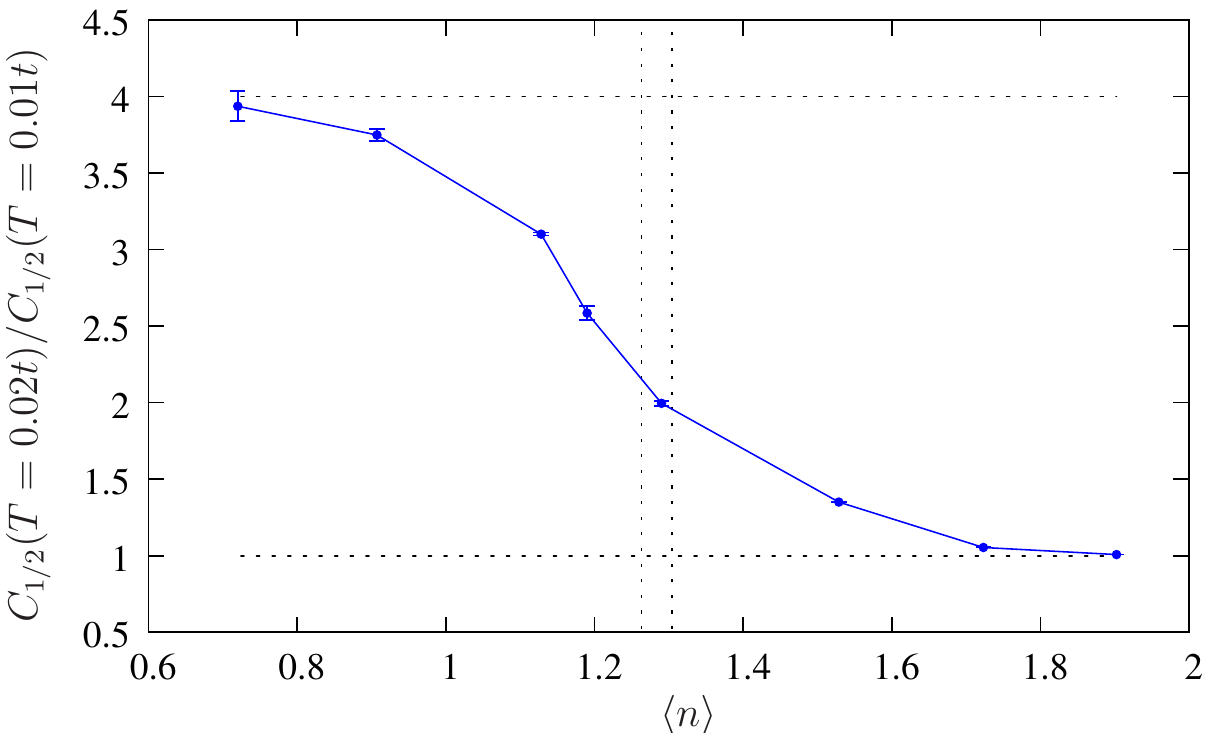} 
\caption{\label{fig:szsz_n} (Color online) Ratio of the spin-spin correlation function $C_{SS}(\tau=1/(2T))$ for temperatures $T=0.02t$ and $0.01t$ as a function of density. Vertical lines mark the approximate location of the transition, where the self-energy is consistent with a $(\nu_{n}/t)^{\alpha}$ behavior with $\alpha\sim 0.5$.
On the Fermi liquid side, obviously $C_{SS}(\tau=1/(2T))\sim T^2$, while $C_{SS}(\tau=1/(2T)) = \rm{const.}$ in the frozen moment phase.
}
\end{figure}

The phasediagram in the plane of filling $\av{n}$ and interaction $U$ is shown in Fig. \ref{fig:phasediagram_sf} for $J/U=1/6$ and temperature $T/t=0.02$. It reproduces the qualitative features reported in Ref. \onlinecite{spinfreezing} for the three-orbital model: For small values of the density and interaction, we find a Fermi liquid phase, while for larger values, the model exhibits a frozen moment phase. The frozen moment phase is characterized by a spin-spin correlation function $C_{SS}(\tau)\Let \av{S_{z}(\tau)S_{z}(0)}$ that approaches a non-zero constant at large times, as shown in Fig. \ref{fig:szsz_tau}. In this phase, the spin-spin correlation function at $\tau=1/(2T)$ (which we refer to as $C_{1/2}$), is expected to be independent of temperature.
In a Fermi liquid at low temperature, on the other hand, the spin-spin correlation function behaves as $C_{SS}(\tau)\sim(T/\sin(\pi\tau T))^{2}$ for times $\tau$ sufficiently far from $\tau=0$ or $1/T$, respectively, so that $C_{1/2}\sim T^{2}$. In Fig. \ref{fig:szsz_n} we show the ratio $C_{1/2}(T=0.02t)/C_{1/2}(T=0.01t)$, as a function of filling, which clearly confirms this behavior. The ratio changes from the value $4$ expected in the Fermi liquid phase to $1$ expected for frozen moments, passing through $2$ near the spin-freezing transition, indicating a $T$-linear behavior. Hence, in the vicinity of the transition, the spin-spin correlation function decays unusually slowly, $C_{SS}(\tau)\sim 1/\tau$, consistent with the findings in Ref.~\onlinecite{spinfreezing}.

We note that a similar phasediagram for a two-orbital model, which plots the ``quasi-particle weight" $Z$ in the space of filling and interaction strength, has recently been reported in Ref.~\onlinecite{Luca11}. Our spin-freezing transition line seems to resemble the contour-lines for fixed $Z$ in Ref.~\onlinecite{Luca11}, although at the temperature $T/t=0.02$ of our calculation, the strong deviations from Fermi-liquid behavior near the transition mean that in this regime a quasi-particle weight cannot be properly defined. It is an interesting open question whether and how the properties of a (strongly renormalized) Fermi-liquid are recovered as $T\rightarrow 0$. 

\begin{figure}[b]
\includegraphics[scale=0.675,angle=0]{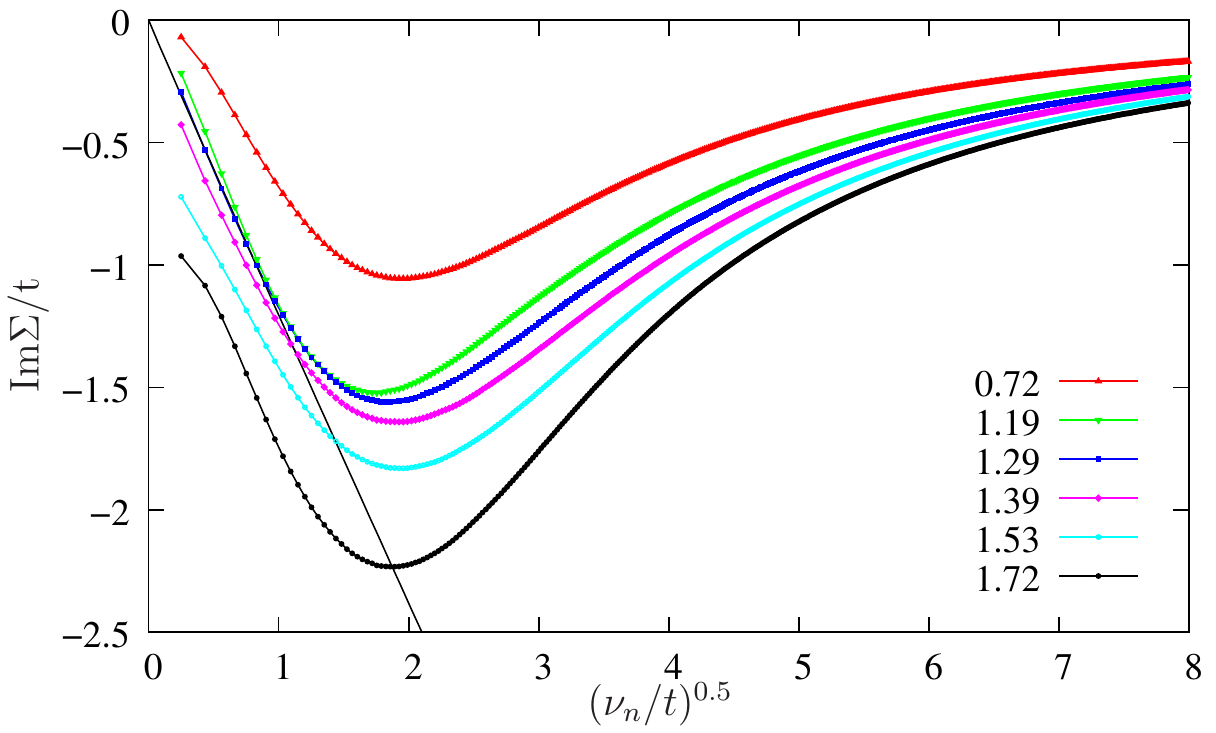} 
\caption{\label{fig:sigma_sf} (Color online) Imaginary part of the self-energy for various fillings for the two-orbital model at $U/t=8$, obtained using the improved estimator measured in the Legendre basis. The positions of the corresponding parameters are marked by circles in the phasediagram in Fig. \ref{fig:phasediagram_sf}. The solid line is a fit proportional to $(\nu_{n}/t)^{\alpha}$, with $\alpha=0.49$.
}
\end{figure}

The development of frozen moments is accompanied by a simultaneous change in the low-frequency behavior of the self-energy. In Fig.~\ref{fig:sigma_sf} we plot the imaginary part of the self-energy as a function of Matsubara frequencies for $U/t=8$ and the densities indicated by circles in the phasediagram in Fig.~\ref{fig:phasediagram_sf}. 
For small densities, the imaginary part of the self-energy $\Im\Sigma(i\nu_{n}\to 0)$ extrapolates to zero as expected for a Fermi liquid. In the presence of frozen moments, however, the electrons are expected to be scattered by these moments, resulting in a non-zero value 
$-\Im\Sigma(i\nu_{n}\to 0)=\Gamma\sgn(\nu_{n})$.
This behavior is evident from the imaginary time data. We find that in the transition region the low-frequency behavior of the self-energy is well described by a power-law $\Im\Sigma(\nu_{n})/t=\gamma(\nu_{n}/t)^{\alpha}$. The phase boundary in Fig.~\ref{fig:phasediagram_sf} corresponds to a ``square-root" frequency dependence, i.~e. $\alpha= 0.5$. Note that the horizontal axis is $(\nu_n/t)^{0.5}$, so that $\alpha=0.5$ corresponds to a linear increase at low frequencies (black line in Fig.~\ref{fig:sigma_sf}).

\begin{figure}[t]
\includegraphics[scale=1,angle=0]{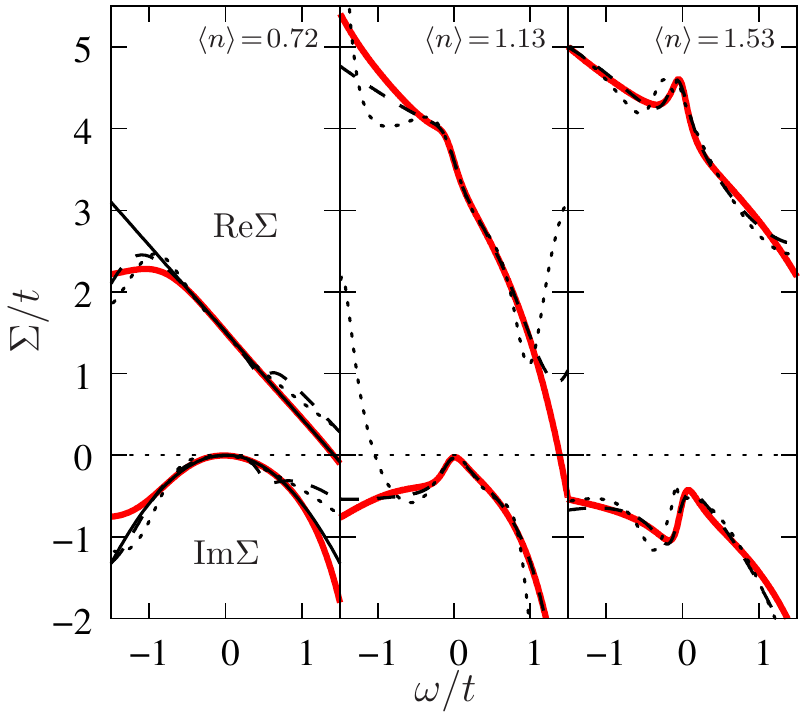} 
\caption{\label{fig:sigmaE} (Color online) Self-energy on the real axis obtained from the Matsubara axis data by analytical continuation using Pad\'e approximants.
Thick lines (red, gray in print) show the result obtained from the improved estimator measured in the Legendre basis. The dashed and dotted lines show results obtained from the improved estimator and Dyson's equation, respectively, measured in Matsubara basis.
The positions of the parameters in the phasediagram are marked by squares in Fig. \ref{fig:phasediagram_sf}.
In the Fermi liquid phase (left panel) the data obtained from the improved estimators measured in the Legendre basis is well described by the Fermi liquid form $\Sigma(\omega)\sim (1-1/Z)\omega - i\mu\omega^{2}$ over a wide energy range (thin solid lines), while the data obtained from the two other methods exhibits spurious features.
}
\end{figure}

It is instructive to compute the self-energy on the real axis.  Analytical continuation is a delicate issue, in particular for Monte Carlo data, due to the presence of statistical noise. Given the high quality of the Matsubara-axis data of Fig.~\ref{fig:sigma_sf}, which have been obtained using the combination of improved estimator and Legendre noise filter, we nevertheless attempt to perform an analytical continuation using Pad\'e approximants.\cite{pade}
The results are shown by thick red (gray) lines in Fig.~\ref{fig:sigmaE}. The three values of the density correspond to the positions marked by squares in the phasediagram in Fig. \ref{fig:phasediagram_sf}.
We first observe that for $\av{n}=0.72$, the real axis self-energy has the expected Fermi liquid form and is well described by the expression $\Sigma(\omega)\sim \kappa + \lambda\omega - i\mu\omega^{2} $ over a wide energy range (thin solid line). 
For comparison we have also plotted the results obtained from the self-energy measured in Matsubara representation, using Dyson's equation (dotted lines) and the improved estimator (dashed). 
In particular the self-energy obtained from Dyson's equation shows spurious features and a departure from the expected Fermi liquid behavior already for energies close to the Fermi level. 

By construction, the Pad\'e analytical continuation yields the most accurate result for small $\omega$. As a consistency check, we performed a least-squares fit of the Pad\'e data to the Fermi liquid form in the energy window $[-0.1:0.1]$  (thin solid line), finding $\mu=1.24$ and $\lambda=(1-1/Z)=-1.0899$, which corresponds to a weakly renormalized Fermi liquid with $Z=0.479$.
The value of $\lambda$ agrees remarkably well with the value obtained from a second-order polynomial extrapolation of the self-energy, $\lambda=-1.0845$ for the improved estimator measured in Legendre ($\lambda=-1.0904$ for the self-energy obtained from Dyson's equation). The deviation is less than 0.5\%, while the fit of the Pad\'e analytical continuation of the self-energy obtained from Dyson's equation already disagrees by more than 7\%. 
We note that the Monte Carlo error on the self-energy is of the order of $2\cdot 10^{-4}$. A summary of the parameters extracted from the Pad\'e analytical continuation for the various measurement procedures is given in Table~\ref{table1}.

\begingroup \squeezetable
\begin{table}[t]
\begin{center}
\begin{tabular}{lcllllll}
\toprule
& & \multicolumn{2}{c}{Leg. + impr. est.} & \multicolumn{2}{c}{impr. est.} & \multicolumn{2}{c}{Dyson}\\ 
$\Delta E$ & \#df & $\lambda$ & $\sigma$ & $\lambda$ & $\sigma$ & $\lambda$ & $\sigma$\\
\hline
0.02 & 0 & $-1.09052$ & -- & $-1.08322$ & -- & $-1.20196$ & --\\
0.1 & 12 & $-1.08994$ & $6.0\!\cdot\! 10^{-5}$& $-1.08667$ & $8.6\!\cdot\! 10^{-5}$ & $-1.16301$ & $1.9\!\cdot\! 10^{-3}$\\
0.25 & 32 & $-1.0857$ & $3.7\!\cdot\! 10^{-4}$& $-1.103$ & $5.4\!\cdot\! 10^{-3}$ & $-1.064$ & $9.3\!\cdot\! 10^{-3}$\\
0.5 & 64 & $-1.070$ & $3.5\!\cdot\! 10^{-3}$& $-1.11$ & $1.4\!\cdot\! 10^{-2}$ & $-1.03$ & $1.2\!\cdot\! 10^{-2}$\\
\hline
\end{tabular}
\caption{\label{table1} Coefficient $\lambda=1-1/Z$ obtained from a least-squares fit of the Pad\'e data in the energy window $[-\Delta E:\Delta E]$ using \#df degrees of freedom. The values should be compared with the result obtained from a second-order polynomial extrapolation of the imaginary part of the Matsubara self-energy, which yields $\lambda=-1.08449$.
}
\end{center}
\end{table}
\endgroup

\begin{figure}[b]
\includegraphics[scale=0.725,angle=0]{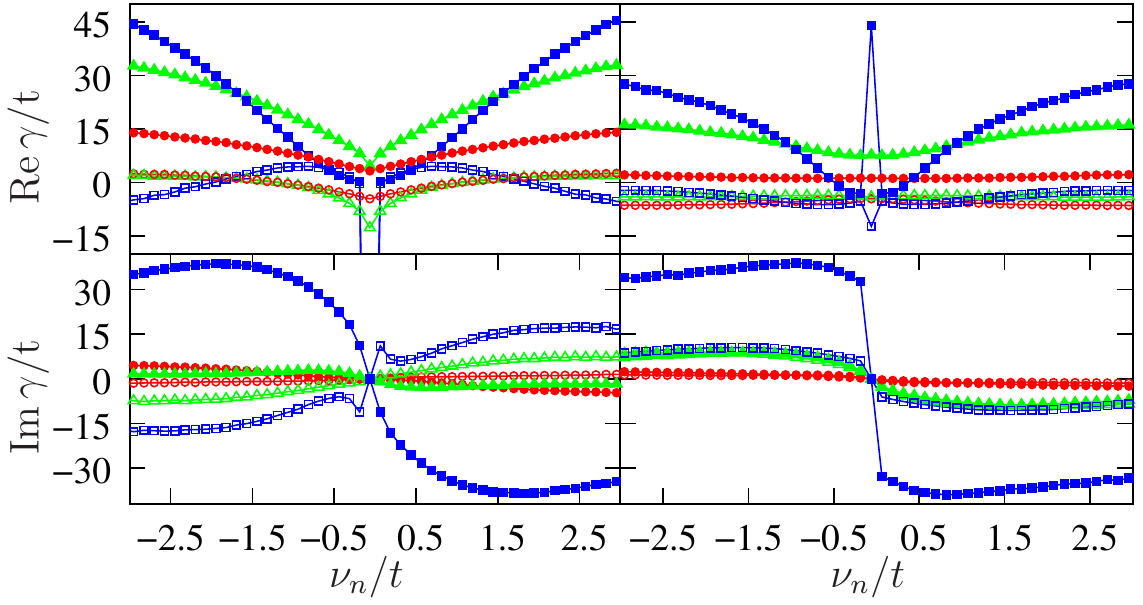} 
\caption{\label{fig:vertex_sf} (Color online) Real and imaginary parts of the reducible vertex function $\gamma_{\nu\nu'\omega}$ for the two-orbital Hubbard model for $\nu'=-\pi T$ and $\omega=2\pi T$ across the spin-freezing transition. The left panels shows intra- and the right panels interorbital components of the spin (closed symbols) and charge (open symbols) channels. Circles (red), triangles (green) and squares (blue) correspond to densities $\av{n}=0.72,1.10$ and $1.72$, respectively. The position of these parameters in the phasediagram is marked by stars in Fig. \ref{fig:phasediagram_sf} (in the corresponding color). The y-axis range in the upper left panel has been restricted to improve visualization.
}
\end{figure}

The self-energy for $\av{n}=1.13$ (middle panel) shows incipient non-Fermi liquid behavior. Both the real and imaginary parts develop a peak around the Fermi level, qualitatively consistent with the appearance of a ``square-root" non-analyticity near the spin-freezing transition.
 While the results from the improved estimators agree fairly well, the result obtained using Dyson's equation deviates rather strongly.
Finally, for $\av{n}=1.53$ the self-energy clearly exhibits non-Fermi liquid behavior, with even more pronounced peaks around the Fermi level. The non-zero value of the imaginary part at zero frequency is again consistent with the value obtained from a polynomial extrapolation of the Matsubara self-energy. 
As a final illustration for this model, we calculate the reducible vertex function according to Eqs. \eqref{eqn:gamma}-\eqref{eqn:chiconnfinal}. Figure~\ref{fig:vertex_sf} illustrates the evolution of the vertex for fixed $\nu'=-\pi T$ and bosonic Matsubara frequency $\omega=2\pi T$ across the spin freezing transition, both for intra-orbital (left panels) and inter-orbital (right panels) components. The vertex is essentially featureless on the Fermi liquid side ($\av{n}=0.72$, red circles), but develops structure  as the spin-freezing transition is approached ($\av{n}=1.10$, green triangles), and notably in the the frozen moment phase ($\av{n}=1.72$, blue squares). In particular the spin-channel (closed symbols) is strongly enhanced in the frozen moment phase.

One can anticipate that these features will induce significant changes when the vertex is used to calculate the momentum dependent self-energy in diagrammatic extensions of dynamical mean-field theory.

\subsection{High-spin to low-spin transition}

\begin{figure}[b]
\includegraphics[scale=0.675,angle=0]{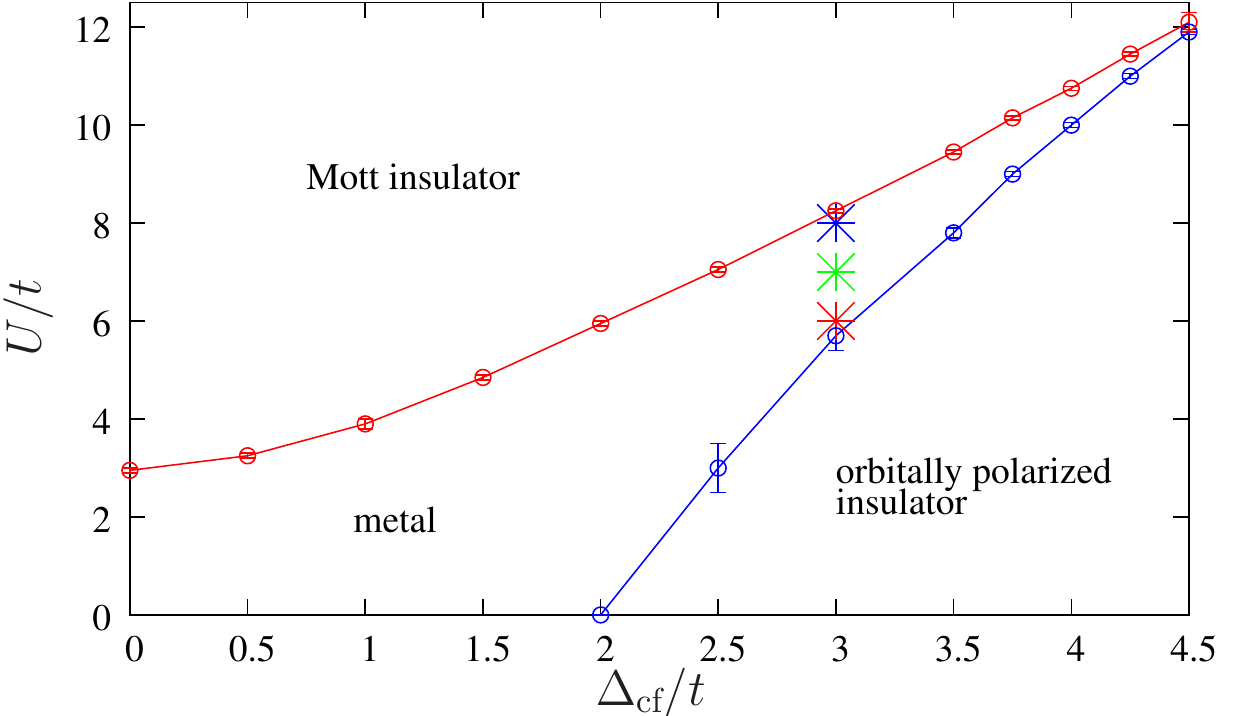} 
\caption{\label{fig:phasediagram_hs} (Color online) Phasediagram in the space of crystal field splitting $\Delta_{\text{cf}}$ and interaction strength $U$ of the half-filled two-orbital Hubbard model on the Bethe lattice with bandwidth $4t$, $J/U=1/4$ and $T/t=0.02$. The parameters for which the vertex function in Fig. \ref{fig:vertex_hs} has been calculated are marked by stars.
}
\end{figure}

\begin{figure}[t]
\includegraphics[scale=0.7,angle=0]{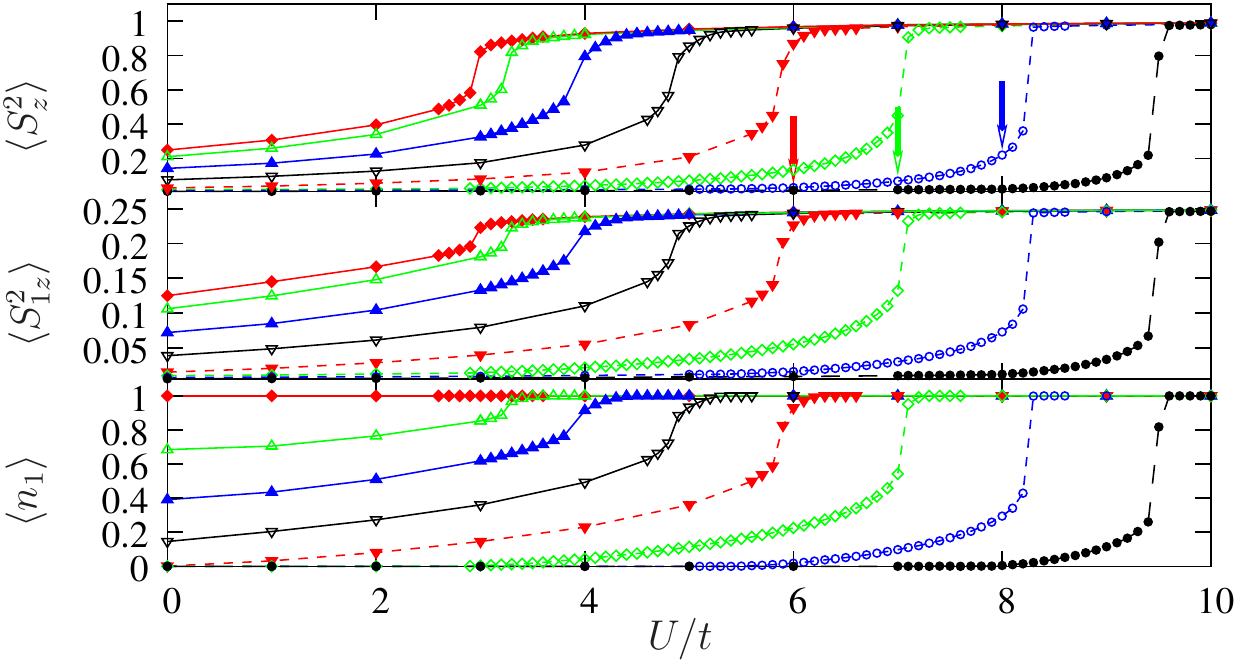} 
\caption{\label{fig:szsz_hs} (Color online) Total moment $\av{S_{z}^{2}}$, moment of orbital 1, $\av{S_{1z}^{2}}$ and density of orbital 1, $\av{n_{1}}$ as a funciton of $U$ for various values of the crystal field splitting $\Delta_{\text{cf}}$. From left to right (and top to bottom) the curves correspond to the values $\Delta_{\text{cf}}=0,0.5,1,1.5,2,2.5,3,3.5$.
}
\end{figure}

\begin{figure}[b]
\includegraphics[scale=0.725,angle=0]{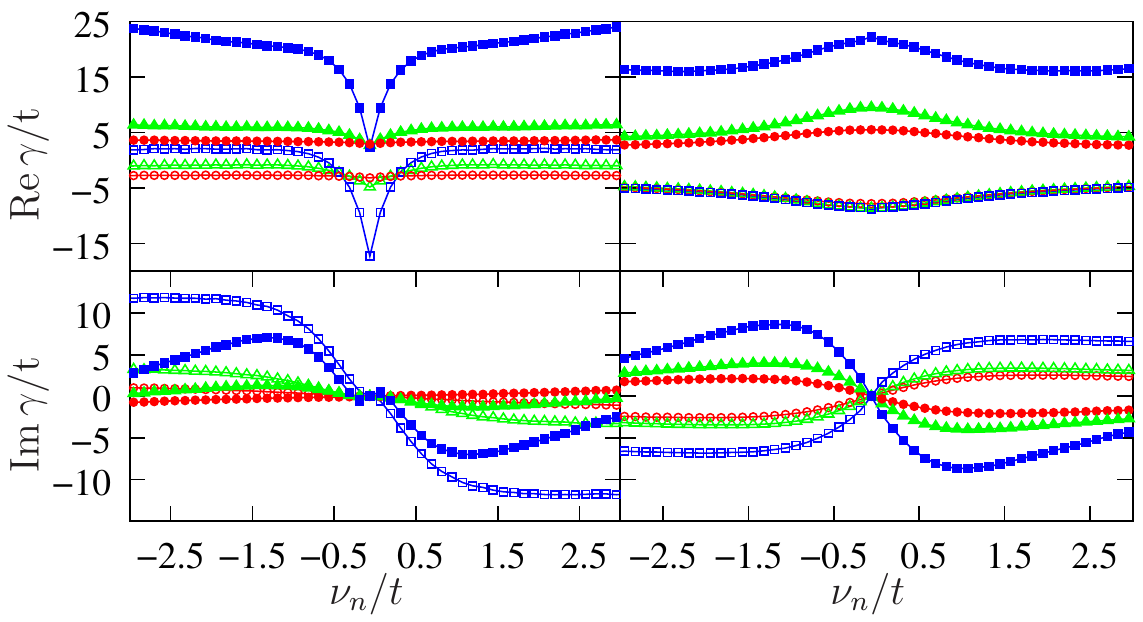} 
\caption{\label{fig:vertex_hs} 
(Color online) Real and imaginary parts of the reducible vertex function $\gamma_{\nu,\nu',\omega}$ for the two-orbital Hubbard model for $\nu'=-\pi T$ and $\omega=2\pi T$ at $\Delta_{cf}/t=3.0$ for the points indicated by stars (in corresponding color) in the phase diagram of the low-spin to high-spin transition in Fig.~\ref{fig:phasediagram_hs}.
The left panel shows intraorbital (for orbital 2) and the right panel interorbital components of the spin (closed symbols) and charge (open symbols) channels. Circles (red), triangles (green) and squares (blue) correspond to $U/t=6.0,7.0$ and $8.0$. The corresponding points in the phasediagram are marked by stars in Fig. \ref{fig:phasediagram_hs} (in corresponding color).
}
\end{figure}

As a second application, we study the same model at half-filling and additionally consider a crystal-field splitting term $\hat{\Delta}_{\text{cf}}$, defined as
\begin{align}
\hat{\Delta}_{\text{cf}} = \sum_{\sigma} \Delta_{\text{cf}} (n_{1,\sigma}-n_{2,\sigma})
\end{align}
in Eq.~\eqref{eqn:hamiltonian}. This model has recently been considered as a minimal model for the physics of LaCoO$_{3}$.\cite{kuneshs2ls}
The condition for half-filling is found by taking the chemical potential $\mu$ to be half the sum over the elements of any row or column of the interaction matrix, which gives $\mu_{1/2}\equiv 3U/2 - 5J/2$.
For this model, a high-spin to low-spin, and associated orbital polarization transition have been reported.\cite{hsp2losp} 
To quantify the effect of our density-density approximation, we compute the phasediagram in the space of crystal field splitting and interaction strength. Comparison of the result shown in Fig.~\ref{fig:phasediagram_hs} with the phasediagram for the rotationally invariant interaction (including spin-flip and pair-hopping terms) reported in Ref.~\onlinecite{hsp2losp} shows that the differences are rather marginal and that all qualitative features are reproduced by the density-density approximation. In particular we find a high-spin Mott insulating phase at large interaction and small crystal field splitting, and a low-spin orbitally polarized insulator at small interaction and large crystal field splitting. These two qualitatively distinct insulating phases are separated by a metallic phase with an end-point near $\Delta_\text{cf}/t=4.5$.

The nature of the two insulating phases becomes evident if one plots the spin expectation value (total moment $\av{S_{z}^{2}}$ or moment of orbital 1 $\av{S_{1z}^{2}}$) and the occupancy of orbital 1 ($\av{n_{1}}$) as a function of interaction strength. Several such traces corresponding to fixed values $\Delta_{\text{cf}}=0,0.5,1,1.5,2,2.5,3,3.5$ are shown in Fig.~\ref{fig:szsz_hs}. In the orbitally polarized insulator, the occupancy of orbital 1 (which is shifted up in energy) is very low. The filling continuously increases with $U$ through the metallic phase and eventually jumps to $\av{n_{1}}\approx 1$ at the transition into the high-spin Mott insulating phase. A similar behavior is seen in the traces for $\av{S_{z}^{2}}$.

We finally show in Fig.~\ref{fig:vertex_hs} the evolution of the reducible vertex function in the metallic phase, as one moves from the phase boundary to the orbitally polarized insulator to the phase boundary with the high-spin Mott insulator. The three data points, corresponding to $\Delta_{\text{cf}}/t=3.0$ and $U/t=6$, $7$ and $8$ are marked in the phasediagram (Fig.~\ref{fig:phasediagram_hs}) by star symbols. While rather featureless near the phase boundary to the orbitally polarized insulator, the vertex develops structure as the moment $\av{S_{z}^{2}}$ increases (cf. the arrows in the upper panel of Fig.~\ref{fig:szsz_hs}).

\section{Conclusions and outlook}

We presented efficient measurement procedures for the self-energy and vertex function within the hybridization expansion CTQMC approach. The improved estimators are based on higher-order correlation functions, which can be particularly easily measured in single-site (multi-orbital) models with density-density interactions, but also for generic models. In combination with a recently proposed noise-filtering scheme (Legendre representation), we were able to completely eliminate the noise problems at intermediate to high frequencies, which have plagued the ``standard" evaluation of the self-energy and vertex function using the hybridization expansion method.
With the noise-problem solved, one can fully exploit the performance advantages of CT-HYB in the strong-correlation regime and in the case of (single-site) multiorbital models. The accurate measurement of two-particle Green's functions and vertex functions should enable dual fermion\cite{dualfermion1,ldfa} or dynamical vertex approximation\cite{toschi} calculations of multi-band systems, and thus (in combination with band-structure input) the \emph{ab-initio} simulation of transition metal compounds which capture the effect of non-local correlations. 
While we have presented results for two-orbital models, the computational effort for the measurement of the vertex function scales as the square of the number of orbitals (for diagonal hybridization) and thus simulations are feasible even for five-orbital models on a small computer cluster.

We have further demonstrated the efficiency of the improved measurements with self-energy and vertex data for a two-orbital model with density-density interactions. In particular, we have revisited the phenomenon of the ``spin freezing" transition, which was originally discovered in a three-orbital model, but which manifests itself in a very analogous manner also in the two-orbital case. In combination with the results by Ishida and Liebsch \cite{Ishida2010} for five-orbital models, this establishes the spin-freezing phenomenon as a generic feature of multi-orbital models with large Hund's rule coupling. This phenomenon leads to strong non-Fermi liquid effects in certain $(U,n)$ regions of the paramagnetic metallic phase. 

\acknowledgments
We would like to thank L.~Boehnke, M.~Eckstein, M.~Ferrero, J.~Mravlje, and O.~Parcollet for valuable discussions. The simulations have been performed on the Brutus cluster at ETH Z\"urich, using an  implementation based on the ALPS-libraries.\cite{ALPS2}

\appendix
\section{Self-energy}
\label{app:sigma}

Following Ref.~\onlinecite{NRG}, we derive the expression for the self-energy for the Hamiltonian \eqref{eqn:hamiltonian}. The single-particle Green's function is defined as
\begin{align}
G_{ab}(\tau-\tau') \Let -\av{\T c_{a}(\tau)c^{\dagger}_{b}(\tau')} .
\end{align}
In order to arrive at the equation of motion, for reasons outlined in appendix \ref{app:vertex}, we take the derivative with respect to its second argument,
\begin{align}
&\dtau_{'}  G_{ab}(\tau-\tau') =\nonumber\\
& \delta(\tau-\tau') \av{c_a(\tau) c_b^\dagger(\tau')} + \delta(\tau-\tau')\av{c_b^\dagger(\tau') c_a(\tau)}\nonumber\\
& - \theta(\tau-\tau')\av{c_a(\tau) \dtau_{'} c_b^\dagger(\tau')} + \theta(\tau'-\tau)\av{\dtau_{'} c_b^\dagger(\tau') c_a(\tau)}\nonumber\\
&= \delta(\tau-\tau') \av{\{c_a(\tau), c_b^\dagger(\tau')\}} - \av{\T c_a(\tau) \dtau_{'} c_b^\dagger(\tau')}\nonumber\\
&= \delta(\tau-\tau')\delta_{ab} - \av{\T c_a(\tau) \dtau_{'} c_b^\dagger(\tau')}.
\label{eqn:app:timederiv}
\end{align}
The equation of motion for the creation operator is
\begin{align}
\dtau_{'} c_b^{\dagger}(\tau') = [H,c_b^{\dagger}](\tau')
\end{align}
and the commutator $[H,c_b^{\dagger}]$ with the Hamiltonian \eqref{eqn:hamiltonian} evaluates to
\begin{align}
& \sum_i \varepsilon_i [n_i,c_b^{\dagger}] + \frac{1}{2}\sum_{ij} U_{ij} [n_{i}n_{j}, c_b^{\dagger}] + \sum_{\kv ij} V_\kv^{*\,ij} [c^\dagger_i f_{\kv j},c_b^{\dagger}]\nonumber\\
&= \varepsilon_b c_b^{\dagger} + \frac{1}{2} \sum_{j} (U_{jb}+ U_{bj}) c_{b}^{\dagger}n_j  + \sum_{\kv j} f_{\kv j}^{\dagger} V_\kv^{*\, jb}.
\label{eqn:commgendag}
\end{align}
Note that the commutator in \eqref{eqn:commgendag} involving the interaction term generates two terms with different order of the operators $c^{\dagger}_{r}$ and $n_{j}$. These can be commuted since the interaction matrix for identical flavors is zero: $U_{jj}=0$. Inserting this expression into \eqref{eqn:app:timederiv}, we obtain
\begin{align}
\dtau_{'} G_{ab}(\tau-\tau') 
=& \delta(\tau-\tau')\delta_{ab} + G_{ab}(\tau-\tau') \varepsilon_b \nonumber\\
&+ \frac{1}{2}\sum_{j} (U_{jb}+U_{bj}) F^{j}_{ab}(\tau-\tau') \nonumber\\
&+ \sum_{\kv j} G_{\kv aj}^\text{cf}(\tau-\tau') V_\kv^{*\,jb},
\label{eqn:app:gtderiv}
\end{align}
where we have used the above definition of Green's function and introduced the functions
\begin{align}
G_{\kv ab}^\text{cf}(\tau-\tau') &\Let -\av{\T c_{a}(\tau) f^\dagger_{\kv b}(\tau')}\nonumber\\
F^{j}_{ab}(\tau-\tau') &\Let -\av{\T c_a(\tau) c^\dagger_b(\tau')n_j(\tau') }.
\end{align}
Using the definition of the Fourier transform
\begin{equation}
G(\iom) = \frac{1}{\beta}\int_0^\beta d\tau \int_0^\beta d\tau'\, G(\tau-\tau') e^{\iom (\tau-\tau')},
\end{equation}
\eqref{eqn:app:gtderiv} can be written in Fourier space as
\begin{align}
G_{ab}(\iom)(\iom - \varepsilon_b) =& \delta_{ab}  + \sum_{\kv j} G_{\kv aj}^\text{cf}(\iom)V_\kv^{*\,jb} \nonumber\\
& + \frac{1}{2}\sum_{j} (U_{jb}+U_{bj}) F^{j}_{ab}(\iom).
\label{eqn:eomfourier}
\end{align}
Note that here no summation over repeated indices is implied (unless explicitly indicated). In DMFT it is customary to define the bare Green's function of the impurity model as
\begin{equation}
G_{0\, ab}^{-1}(\iom) = (\iom -\varepsilon_b)\delta_{ab} - \Delta_{ab}(\iom),
\end{equation}
where $\Delta_{ab}(\iom)$ is the hybridization function. 
Subtracting $\sum_{i}G_{ai}(\iom)\Delta_{ib}(\iom)$ on both sides of \eqref{eqn:eomfourier}, we can rewrite its left hand side as
\begin{align}
\sum_{i}G_{ai}(\iom)\left[(\iom-\varepsilon_{b})\delta_{ib}-\Delta_{ib}(\iom)\right]=\sum_{i}G_{ai}(\iom)G_{0,ib}^{-1}.
\end{align}
Hence multiplying both sides of \eqref{eqn:eomfourier} by $G_{0}$ from the right using matrix multiplication, the equation of motion becomes
\begin{align}
G_{ab}(\iom) = G_{0, ab} &- \sum_{ij}G_{ai}(\iom)\Delta_{ij}(\iom)G_{0,jb}(\iom)\nonumber\\
&+\sum_{\kv ij} G_{\kv ai}^\text{fc}(\iom)  V_\kv^{*\,ij} G_{0\, jb}(\iom)\nonumber\\
&+ \frac{1}{2}\sum_{ij} (U_{ji}+U_{ij}) F^{j}_{ai}(\iom)G_{0\, ib}(\iom).
\label{eqn:eomfourier2}
\end{align}
The Green's function $G_{\kv ab}^\text{fc}(\tau-\tau')$ in turn is determined from its equation of motion,
\begin{align}
\dtau_{'} G_{\kv ab}^\text{cf}(\tau-\tau') = -\av{\T c_{a}(\tau) \dtau_{'} f^\dagger_{\kv b}(\tau')},
\label{eqn:eomGcf}
\end{align}
where now
\begin{align}
\dtau_{'} f_{\kv b}^{\dagger}(\tau') = [H, f_{\kv b}^{\dagger}](\tau').
\end{align}
The commutator is
\begin{align}
[H,f_{\kv b}^{\dagger}] &= \sum_{\kv' i} \varepsilon_{\kv'}^i [n_{\kv' i}, f_{\kv b}^{\dagger}] + \sum_{\kv' ij} V_{\kv'}^{ij} [c^{\dagger}_{i}f_{\kv' j}, f^{\dagger}_{\kv b}]\nonumber\\
&=\varepsilon_\kv^b f^{\dagger}_{\kv b} + \sum_j  c_j^{\dagger} V_\kv^{jb}.
\end{align}
Inserting this back into \eqref{eqn:eomGcf} yields
\begin{align}
\dtau_{'} G_{\kv ab}^\text{cf}(\tau-\tau') = G_{\kv ab}^\text{cf}(\tau-\tau')\varepsilon_\kv^b + \sum_i G_{ai}(\tau-\tau') V_\kv^{ib}.
\end{align}
In Fourier space, this becomes
\begin{equation}
G_{\kv ab}^\text{cf}(\iom) = \sum_i G_{ai}(\iom)\frac{V_\kv^{ib}}{\iom - \varepsilon_\kv^{b}}.
\label{eqn:gcffourier}
\end{equation}
As can be shown by integrating out the bath fermions from the Hamiltonian \eqref{eqn:hamiltonian}, the hybridization function for this model reads
\begin{align}
\Delta_{ab}(\iom) = \sum_{\kv j}\frac{V_\kv^{aj} V_\kv^{* jb}}{\iom-\varepsilon_\kv^j}.
\end{align}
Hence by inserting \eqref{eqn:gcffourier} into \eqref{eqn:eomfourier2} on sees that the terms involving the hybridization function cancel exactly:
\begin{align}
G_{ab}(\iom) 
&= G_{0, ab} + \frac{1}{2}\sum_{ij} (U_{ij}+U_{ji}) F^{i}_{aj}(\iom) G_{0\, jb}(\iom).
\end{align}
Comparing this with Dyson's equation,
\begin{equation}
G_{ab}(\iom) = G_{0\, ab}(\iom) + \sum_{ij} G_{ai}(\iom) \Sigma_{ij}(\iom) G_{0\, jb}(\iom) ,
\end{equation}
we can finally identify the self-energy as
\begin{equation}
\Sigma_{ab}(\iom) = \frac{1}{2}\sum_{ij} G^{-1}_{ai}(\iom) (U_{jb}+U_{bj}) F^{j}_{ib}(\iom) .
\label{eqn:app:sigmafinal}
\end{equation}

\section{Vertex function}
\label{app:vertex}

In order to derive an analogous expression for the connected part of the two-particle Green's function or the impurity vertex function, respectively, we make use of the following operator identity (see e.g. Ref. \onlinecite{2ptmatrix}):
\begin{align}
\partial_{\tau_{a}}& \av{\T A(\tau_a)B(\tau_b)C(\tau_c)D(\tau_d)} =\nonumber\\
&\av{\T \partial_{\tau_{a}} A(\tau_a)B(\tau_b)C(\tau_c)D(\tau_d)}\nonumber\\
+&\delta(\tau_a-\tau_b)\av{\T [A(\tau_a),B(\tau_a)]_\pm C(\tau_c)D(\tau_d)}\nonumber\\
+&\delta(\tau_a-\tau_c)\av{\T [A(\tau_a),C(\tau_a)]_\pm D(\tau_d)B(\tau_b)}\nonumber\\
+&\delta(\tau_a-\tau_d)\av{\T [A(\tau_a),D(\tau_a)]_\pm B(\tau_b)C(\tau_c)} ,
\label{eqn:identity}
\end{align}
where $[A,B]_\pm$ is the (anti-)commutator $[A,B]_\pm\Let AB\pm BA$ depending on whether one deals with fermionic or bosonic operators.
For the two-particle Green's function defined as
\begin{align}
\chi_{abcd}(\tau_a,\tau_b,\tau_c,\tau_d) \Let  \av{\T c_a(\tau_a) c_b^\dagger(\tau_b) c_c(\tau_c) c_d^\dagger(\tau_d)},
\end{align}
we have
\begin{align}
\partial_{\tau_{a}} \chi_{abcd}(\tau_a,\tau_b,\tau_c,\tau_d) =& \partial_{\tau_{a}} \av{\T c_a(\tau_a) c_b^\dagger(\tau_b) c_c(\tau_c) c_d^\dagger(\tau_d)}\nonumber\\
=& \av{\T \partial_{\tau_{a}} c_a(\tau_a) c_b^\dagger(\tau_b) c_c(\tau_c) c_d^\dagger(\tau_d)}\nonumber\\
&- \delta(\tau_a-\tau_b) G_{cd}(\tau_c-\tau_d) \delta_{ab}\nonumber\\
&+ \delta(\tau_a-\tau_d) G_{cb}(\tau_c-\tau_b) \delta_{ad} .
\end{align}
The last two terms stem from the discontinuities of $\chi_{abcd}(\tau_a,\tau_b,\tau_c,\tau_d)$ at $\tau_{a}=\tau_{b}$ and $\tau_{a}=\tau_{d}$. 
Defining the Fourier transform
\begin{widetext}
\begin{equation}
\mathcal{F}[f(\tau_a,\tau_b,\tau_c,\tau_d)] \Let \frac{1}{\beta}\int_0^\beta d\tau_a \int_0^\beta d\tau_b \int_0^\beta d\tau_c \int_0^\beta d\tau_d\, f(\tau_a,\tau_b,\tau_c,\tau_d) e^{\iom_a\tau_a}e^{-\iom_b\tau_b}e^{\iom_c\tau_c}e^{-\iom_d\tau_d} \equiv f(\iom_a,\iom_b,\iom_c,\iom_d),
\end{equation}
\end{widetext}
one has
\begin{align}
\mathcal{F}[\delta(\tau_a-\tau_b)G_{cd}(\tau_c-\tau_d)] &= \beta\delta_{\nu_a,\nu_b}G_{cd}(\iom_c) \delta_{\nu_c,\nu_d},\nonumber\\
\mathcal{F}[\chi_{abcd}(\tau_a,\tau_b,\tau_c,\tau_d)] &= \chi_{abcd}(\iom_a,\iom_b,\iom_c,\iom_d).
\end{align}
Using the equation of motion for $c_{a}(\tau_{a})$, i.e. $\partial_{\tau_{a}}c_{a}(\tau_{a})=[H,c_{a}](\tau_{a})$, we need the commutator
\begin{align}
[H, c_a]= -\varepsilon_{a} c_a  -\frac{1}{2} \sum_j (U_{ja}+U_{aj}) n_j c_a - \sum_{\kv j} V_\kv^{aj} f_{\kv j} ,
\end{align}
which leads to
\begin{align}
&(\iom_a-\varepsilon_{a})\chi_{abcd}(\iom_a,\iom_b,\iom_c,\iom_d) = \nonumber\\
&\beta \left(G_{cd}(\iom_c)\delta_{\nu_a,\nu_b}\delta_{\nu_c,\nu_d}\delta_{ab}\right.
 - \left.G_{cb}(\iom_{c})\delta_{\iom_a,\iom_d}\delta_{\iom_c,\iom_b}\delta_{ad}\right)\nonumber\\
&+\sum_{\kv j} V_\kv^{aj}\, \chi_{\kv jbcd}^\text{fccc}(\iom_a,\iom_b,\iom_c,\iom_d)\nonumber\\
&+\frac{1}{2}\sum_j (U_{ja}+U_{aj}) H^j_{abcd}(\iom_a,\iom_b,\iom_c,\iom_d) .
\label{eqn::eqnofmotionchi1}
\end{align}
Again, no summation over repeated indices is implied unless explicitly indicated.
Here we have introduced the correlation functions
\begin{align}
\chi_{\kv abcd}^\text{fccc}(\tau_a,\tau_b,\tau_c,\tau_d) &\Let \av{\T f_{\kv a}(\tau_a) c_b^\dagger(\tau_b) c_c(\tau_c) c_d^\dagger(\tau_d)},\nonumber\\
H_{abcd}^j(\tau_a,\tau_b,\tau_c,\tau_d) &\Let \av{\T n_j(\tau_a) c_{a}(\tau_a) c_b^\dagger(\tau_b) c_c(\tau_c) c_d^\dagger(\tau_d)}.\nonumber\\
\end{align}
The first one can again be eliminated by considering its equation of motion,
\begin{align}
&\partial_{\tau_a} \chi_{\kv abcd}^\text{fccc}(\tau_a,\tau_b,\tau_c,\tau_d) \equiv \nonumber\\
&\qquad \av{\T \partial_{\tau_{a}} f_{\kv a}(\tau_a) c_b^\dagger(\tau_b) c_c(\tau_c) c_d^\dagger(\tau_d)}.
\end{align}
Using the equation of motion for $f_{\kv a}(\tau_{a})$,
\begin{equation}
\partial_{\tau_{a}} f_{\kv a}(\tau_{a}) = -\epsilon_\kv^a f_{\kv a} (\tau_a) - \sum_{j} V_\kv^{* aj} c_j(\tau_a),
\end{equation}
one finds
\begin{align}
\partial_{\tau_{a} }\chi_{\kv abcd}^\text{fccc}(\tau_a,\tau_b,\tau_c,\tau_d) =& 
-\epsilon_{\kv}^a\, \chi_{\kv abcd}^\text{fccc}(\tau_a,\tau_b,\tau_c,\tau_d)\nonumber\\ &- \sum_{j}V_\kv^{* aj}\chi_{jbcd}(\tau_a,\tau_b,\tau_c,\tau_d),
\end{align}
or in Fourier space
\begin{align}
&\chi_{\kv abcd}^\text{fccc}(\iom_a,\iom_b,\iom_c,\iom_d) =\nonumber\\ 
&\qquad\sum_{j}\frac{V_\kv^{* aj}}{\iom_a -\epsilon_\kv^a} \chi_{jbcd}(\iom_a,\iom_b,\iom_c,\iom_d).
\end{align}
Using this result in \eqref{eqn::eqnofmotionchi1} finally yields
\begin{align}
&\sum_{j}[\iom_a-\varepsilon_{a}-\Delta_{aj}(\iom_a)] \chi_{jbcd}(\iom_a,\iom_b,\iom_c,\iom_d) =\nonumber\\
& \beta [G_{cd}(\iom_c)\delta_{\nu_a,\nu_b}\delta_{\nu_c,\nu_d}\delta_{ab} - G_{cb}(\iom_{c})\delta_{\nu_a,\nu_d}\delta_{\nu_c,\nu_b}\delta_{ad}]\nonumber\\
&+\frac{1}{2}\sum_j (U_{ja}+U_{aj}) H^j_{abcd}(\iom_a,\iom_b,\iom_c,\iom_d).
\end{align}
Subtraction of $\sum_{j}\Sigma_{aj}(\iom_{a})\chi_{jbcd}(\iom_{a},\iom_{b},\iom_{c},\iom_{d})$ on both sides of this equation and using the definition of Green's function, $G_{ab}(\iom_{a})\Let [(\iom_{a}-\varepsilon_{a})\delta_{ab}-\Delta_{ab}(\iom_{a})-\Sigma_{ab}(\iom_{a})]^{-1}$, we have on the left-hand-side of this equation $
\sum_{j}G^{-1}_{aj}(\iom_{a}) \chi_{jbcd}(\iom_a,\iom_b,\iom_c,\iom_d)$. Hence multiplying both sides of this equation by $G_{ab}(\iom_{a})$ using matrix multiplication and identifying the disconnected part of the two-particle Green's function
\begin{align}
\chi^{0}_{abcd}(\iom_a,\iom_b,\iom_c,\iom_d) =& \beta [G_{ab}(\iom_a) G_{cd}(\iom_c)\delta_{\nu_a,\nu_b}\delta_{\nu_c,\nu_d}\nonumber\\
& - G_{ad}(\iom_{a})G_{cb}(\iom_{c}) \delta_{\nu_a,\nu_d}\delta_{\nu_c,\nu_b}],
\end{align}
we find for the connected part
\begin{align}
&\chi_{abcd}(\iom_a,\iom_b,\iom_c,\iom_d) - \chi^{0}_{abcd}(\iom_a,\iom_b,\iom_c,\iom_d) = \nonumber\\
&- \sum_{ij}G_{ai}(\iom_a)\Sigma_{ij}(\iom_a)\chi_{jbcd}(\iom_a,\iom_b,\iom_c,\iom_d)\nonumber\\
&+\frac{1}{2}\sum_{ij} G_{ai}(\iom_a) (U_{ji}+U_{ij}) H^j_{ibcd}(\iom_a,\iom_b,\iom_c,\iom_d).
\end{align}
Or, using \eqref{eqn:app:sigmafinal},
\begin{align}
&\chi_{abcd}(\iom_a,\iom_b,\iom_c,\iom_d) - \chi^{0}_{abcd}(\iom_a,\iom_b,\iom_c,\iom_d) = \nonumber\\
&-\frac{1}{2}\sum_{ji} (U_{ij}+U_{ji})F^i_{aj}(\iom_a)\, \chi_{jbcd}(\iom_a,\iom_b,\iom_c,\iom_d)\nonumber\\
&+\frac{1}{2}\sum_{ij} G_{ai}(\iom_a)(U_{ji}+U_{ij}) H^j_{ibcd}(\iom_a,\iom_b,\iom_c,\iom_d).
\end{align}
Note that the inverse Green's function in \eqref{eqn:app:sigmafinal} cancels. If we had used the equation of motion for Green's function taken with respect to its first argument, this would not have been the case (since in this case the inverse Green's function appears to the right of \eqref{eqn:app:sigmafinal}).

\section{Non-density-density Hamiltonians}
\label{eqn:app:gencorr}

For the case of a general, i.e. non-density-density type of interaction
\begin{equation}
H_{U} = \sum_{ijkl} U_{ijkl} c_{i}^{\dagger} c_{j}^{\dagger} c_{k}c_{l}
\end{equation}
the equation of motion involves the commutator
\begin{align}
[c_{i}^{\dagger} c_{j}^{\dagger} c_{k}c_{l},c_{b}^{\dagger}] = c_{i}^{\dagger} c_{j}^{\dagger} c_{k} \delta_{lb} - c_{i}^{\dagger} c_{j}^{\dagger} c_{l} \delta_{kb}.
\end{align}
According to \eqref{eqn:app:timederiv}, the equation of motion in this case generates terms which require one to measure correlation functions of the form
\begin{align}
\av{\T c_{a}(\tau)c_{i}^{\dagger}(\tau') c_{j}^{\dagger}(\tau') c_{k}(\tau')}.
\end{align}

\bibliography{vertex}

\end{document}